\documentclass[sn-mathphys-num,iicol]{sn-jnl}

\usepackage{graphicx}%
\usepackage{multirow}%
\usepackage{amsmath,amssymb,amsfonts}%
\usepackage{amsthm}%
\usepackage{mathrsfs}%
\usepackage[title]{appendix}%
\usepackage{xcolor}%
\usepackage{textcomp}%
\usepackage{manyfoot}%
\usepackage{booktabs}%
\usepackage{algorithm}%
\usepackage{algorithmicx}%
\usepackage{algpseudocode}%
\usepackage{listings}%

\begin{document}

\title[Emergent collective behavior of cohesive, aligning particles]{Emergent collective behavior of cohesive, aligning particles}


\author*[1]{\fnm{Jeanine} \sur{Shea}}\email{j.shea@tu-berlin.de}

\author[1]{\fnm{Holger} \sur{Stark}}\email{holger.stark@tu-berlin.de}

\affil*[1]{\orgdiv{Institut f\"{u}r Theoretische Physik}, \orgname{Technische Universit\"{a}t Berlin}, \orgaddress{\street{Hardenbergstr. 36}, \city{Berlin}, \postcode{10623}, \country{Germany}}}


\abstract{Collective behavior is all around us, from flocks of birds to schools of fish. These systems are immensely complex, which makes it pertinent to study their behavior through minimal models. We introduce such a minimal model for cohesive and aligning self-propelled particles in which group cohesion is established through additive, non-reciprocal torques. These torques cause constituents to effectively turn towards one another. We additionally incorporate an alignment torque, which competes with the cohesive torque in the same spatial range. By changing the strength and range of these torque interactions, we uncover six states which we distinguish via their static and dynamic properties: a disperse state, a multiple worm state, a line state, a persistent worm state, a rotary worm state, and an aster state. Their occurrence strongly depends on initial conditions and stochasticity, so the model exhibits multistabilities. A number of the states exhibit collective dynamics which are reminiscent of those seen in nature.}




\maketitle

\section{Introduction}
\label{sec:intro}

Collective behavior is omnipresent in our lives, from the microscale, with examples such as bacterial colonies~\cite{bacteria_patterns,bacteria_vicsek,bacteria3,bacteria4} and morphogenesis~\cite{zebrafish,tissue,morpho}, to the macroscale, with examples such as flocks of birds~\cite{birds1,birdworm1,pigeons,birds2} and schools of fish~\cite{fish1,herring,fish3}. 
Although systems exhibiting collective behavior are commonplace, they are not well understood due to their immense complexity. In order to improve our understanding of these complex systems, minimal models are often used. Such models attempt to qualitatively replicate collective behavior with only a few simple rules of interaction and thus help to isolate the fundamental aspects governing collective behavior.

One of the earliest models for collective behavior is the Boid model~\cite{boid_model}. In this model, individuals known as `boids' are subject to interaction rules which promote collision avoidance, alignment, and cohesion. Most subsequent collective behavior models use some variation of these three fundamental rules, with the possible addition of further interaction rules. For example, in the behavioral zonal model~\cite{COUZIN20021,Couzin_2005}, these three rules are implemented in different spatial areas around each particle. Depending on the parameters of the different interactions, this model can exhibit swarming, milling, or parallel group motion.  

Although the aforementioned models use both alignment and cohesion to induce collective behavior, it is also possible to achieve collective motion using only alignment \cite{vicsek,ROMANCZUK,Pagonabarraga} or cohesion \cite{Loewen} interactions. Most famously, in the Vicsek model~\cite{vicsek}, particles move together simply by adapting their orientation to align with those of their neighbors. The Vicsek model does not, however, lead to the formation of groups which stay together. In order to achieve this, the alignment interaction must be supplemented by some additional element, such as a vision cone~\cite{Peruani_attraction} or cohesion interactions~\cite{vicsek+cohesion}. 

Oftentimes, in systems where collective behavior is observed, interactions among constituents are non-reciprocal, meaning that Newton's third law is violated and action-reaction symmetry does not hold~\cite{statmech_nr}. Indeed, a diverse range of collective behaviors have been seen for systems in which non-reciprocal interactions are present~\cite{Golestanian_nr,sabine_nr,Kreienkamp_2022,Milos_Nonreciprocal,Alert_TurnAway,Nilsson_2017,Alert_TurnTowards,Fruchart2021NonreciprocalPT,Marchetti_nr,Saha_2019}.

Non-reciprocity can be introduced to a system in many different ways including non-reciprocal force interactions~\cite{Rosalba,NR_force}, interactions in which a vision cone is used~\cite{Peruani_attraction,COUZIN20021,Couzin_2005,Gompper_similar}, and non-reciprocal torque interactions \cite{Nilsson_2017,Alert_TurnTowards,Milos_Nonreciprocal,Alert_TurnAway}. Here, we specifically focus on non-reciprocity via torque interactions. In previous studies of collective behavior induced by non-reciprocal torque interactions, it has been shown that effectively attractive non-reciprocal torques can produce active phase separation \cite{Nilsson_2017,Alert_TurnTowards}. Models with effectively repulsive non-reciprocal torques \cite{Milos_Nonreciprocal,Alert_TurnAway} can exhibit flocking behavior or active phase separation, depending on the exact parameters used.

Here, we introduce a model for cohesive and aligning self-propelled particles in which group cohesion is established through additive, non-reciprocal torques. This model combines elements from the models of Couzin \textit{et al.}~\cite{COUZIN20021,Couzin_2005} and Negi \textit{et al.}~\cite{Gompper_similar}. We use the additive cohesion and alignment torques of Couzin \textit{et al.} in combination with the hard-core potentials of Negi \textit{et al.}. Unlike both of these models, we do not use a vision cone and we keep the interaction range the same for both cohesive and aligning torques. 

In this paper, we explore the different states which emerge from this model when varying the interaction radius and the strength of the torque interactions. We uncover a rich variety of states: a disperse state, a multiple worm state, a line state, a persistent worm state, a rotary worm state, and an aster state. Although the changes to the previous models of Couzin \textit{et al.} and Negi \textit{et al.} are subtle, these changes introduce new dynamics
including in the formation of the different states.
In particular, we observe multistabilities across many states, a 
seemingly generic trait which was not observed
or discussed in either of the aforementioned models. We note that the model of Couzin \textit{et al.} exhibits hysteresis; however, it does not exhibit multistabilities emerging from random initializations, as we observe in our model. Furthermore, the rotary worm state exhibited by our model has not been seen in previous models.

We begin by introducing our model in Section~\ref{sec:syssim}. In Sections~\ref{sec:stat_props} and \ref{sec:dyn_props}, we classify the different observed collective behaviors into distinct states via their static and dynamic properties. We then analyze the resultant state diagram for our chosen range of parameters in Section~\ref{sec:state_dia}. In Section~\ref{sec:worm_props}, we explore in greater depth the persistent and rotary worm states by analyzing their structural properties as well as the dynamics of individual constituents. We summarize and conclude in Section~\ref{sec:conc_out}.

\section{Model and Simulation Details}
\label{sec:syssim}
We consider a 2D system of overdamped active Brownian particles (ABPs) with diameter $\sigma$ which interact via both forces and torques. The equations of motion for the ABPs are:
\begin{align}
\label{eq:abp}
\dot{\mathbf{r}}_i(t)&=v_0\mathbf{u}_i(t)+\mu\sum_{j\neq i}\mathbf{F}_{ij}+\sqrt{2D}~\boldsymbol{\xi}_i(t),\\
\label{eq:abp_rot}
\dot{\phi}_i(t)&=\mu_R\sum_{j\neq i}\mathcal{T}_{ij} +\sqrt{2D_R}~\eta_i(t),
\end{align}
where $v_0$ is the propulsion velocity and $\mathbf{u}_i(t)=(\cos{\phi_i},\sin{\phi_i})$ is {the orientation of particle $i$. Particles have translational and rotational mobilities $\mu$ and $\mu_R$ respectively, such that their diffusive motion can be characterized by translational diffusion coefficient $D=\mu k_\mathrm{B}T$ and rotational diffusion coefficient $D_R=\mu_R k_\mathrm{B}T$, where $T$ is the temperature. The translational and rotational mobilities are related by $\mu/\mu_R=\sigma^2/3$. The variables $\boldsymbol{\xi}_i(t)$ and $\eta_i(t)$ describe delta-correlated noise with zero mean and unit variance.

Particles exert steric forces on each other according to $\mathbf{F}_{ij}=-\nabla_{\mathbf{r}_i}U_{\epsilon,\sigma}(\mathbf{r}_i-\mathbf{r}_j)$. These interactions take the form of truncated and shifted Lennard-Jones potentials with the energy scale $\epsilon/k_\mathrm{B}T=100$ and are cut off at the distance $r_{\mathrm{c}}=2^{\frac{1}{6}}\sigma$ of the potential minimum. This results in purely repulsive interactions according to the Weeks-Chandler-Anderson (WCA) potential~\cite{WCA}: 
\begin{equation}
\label{eq:pot}
U_{\epsilon,\sigma}(\mathbf{r}_i) = 
\begin{cases}
4\epsilon \left( \left( \frac{\sigma}{|\mathbf{r}|}\right)^{12}- \left( \frac{\sigma}{|\mathbf{r}|}\right)^{6}\right)+ \epsilon & |\mathbf{r}|\leq r_c \\
0 & |\mathbf{r}|> r_c.
 \end{cases}
\end{equation}

The particles additionally experience pairwise orientational interactions which promote alignment and cohesion, as we illustrate in Fig.~\ref{fig:interactions}. We use the alignment interaction:
\begin{equation}
\label{eq:prey-predy_align}
\mathcal{T}^A_{ij} = \mathcal{T}_A~\Theta(R-|\mathbf{r}_{ij}|)~\sin{(\phi_j-\phi_i)}, 
\end{equation}
which causes particle $i$ to turn so that it aligns parallel to particle $j$, and the cohesive interaction:
\begin{equation}
\label{eq:prey-prey_group}
\mathcal{T}^C_{ij} = -\mathcal{T}_C~\Theta(R-|\mathbf{r}_{ij}|)~\sin{\left(\theta_{\mathbf{u}_{i}, \mathbf{\hat{r}}_{ij}}\right)},
\end{equation}
which causes particle $i$ to turn towards particle $j$. Here, $\mathbf{r}_{ij} = \mathbf{r}_i-\mathbf{r}_j$ and $\theta_{\mathbf{u}_{i}, \mathbf{\hat{r}}_{ij}}$ is the angle from $\mathbf{u}_{i}$ to $\mathbf{\hat{r}}_{ij}=\frac{\mathbf{r}_{ij}}{|\mathbf{r}_{ij}|}$ (see Fig.~\ref{fig:interactions}). $\mathcal{T}_A$ and $\mathcal{T}_{C}$ are constants which represent the strength of the alignment and cohesion torques respectively. Schematics showing these two interactions are shown in Fig.~\ref{fig:interactions}. We always keep the ratio $\mathcal{T}_A/\mathcal{T}_{C}=2$ constant. We choose this ratio to avoid creating clusters which coalesce, but fail to move in the same direction, which we saw when using ratios of $\mathcal{T}_A/\mathcal{T}_{C}<1$. $R$ is the range of both the alignment and cohesion interactions. Given that both interactions span the same spatial range, these torques are always competing with one another.

Combining these interactions, the total torque interaction is:
\begin{equation}
\label{eq:prey-prey_aligngroup}
\mathcal{T}_{ij} = \mathcal{T}_0~\Theta(R-|\mathbf{r}_{ij}|)\left(\sin{(\phi_j-\phi_i)}-\frac{1}{2}\sin{\left(\theta_{\mathbf{u}_{i}, \mathbf{\hat{r}}_{ij}}\right)}\right),
\end{equation}
where we have set $\mathcal{T}_0=\mathcal{T}_A=2\mathcal{T}_{C}$. Note that, although the alignment interaction obeys Newton's third law, the cohesive interaction is non-reciprocal; therefore, cohesive interaction torques are not necessarily equal and opposite.

\begin{figure} 
  \centering
  \includegraphics[width=\linewidth]{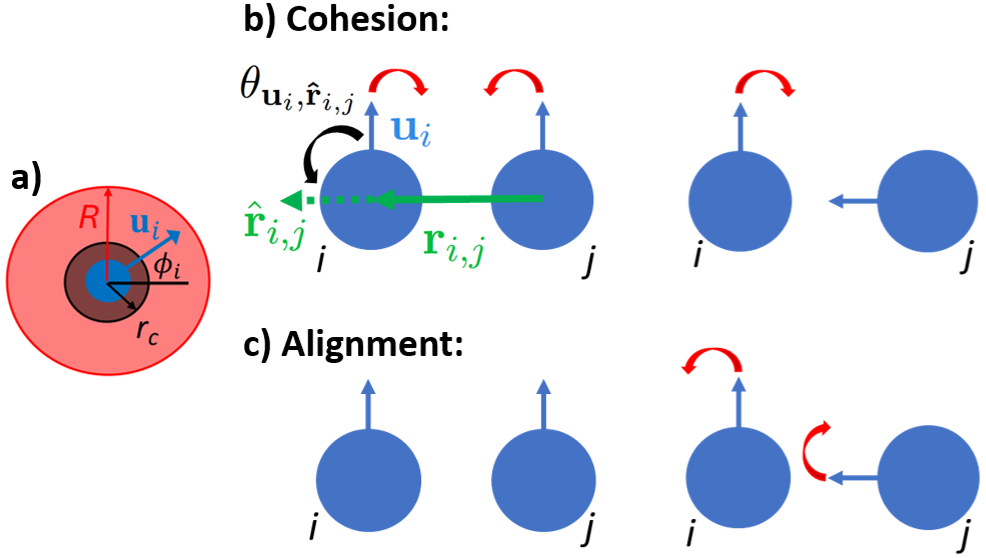}
\caption{Schematic showing a) interaction range $R$ and distance $r_c$ for a given particle $i$ with orientation $\mathbf{u}_i(t)=(\cos{\phi_i},\sin{\phi_i})$, b) cohesive interaction torques between particles $i$ and $j$ (see Eq.~\eqref{eq:prey-prey_group}), and c) alignment interaction torques between particles $i$ and $j$ (see Eq.~\eqref{eq:prey-predy_align}).}
\label{fig:interactions}
\end{figure}

We measure length in units of $\sigma$, energy in units of $k_\mathrm{B}T$, and time in units of $\tau_R\equiv D_R^{-1}$, the active particle reorientation time. Our square simulation box has side lengths $L$ and area $L^2=A$ with periodic boundary conditions in both dimensions. The active bath can be characterized by the packing fraction, $\Phi=N\sigma^2\pi/(4A)$, and the Peclet number, $\mathrm{Pe}=\sigma v_0/D$. For our simulations, we use a Peclet number $\mathrm{Pe}=80$ and a packing fraction of $\Phi=0.025$ for $N=1000$ particles. 

We simulate four runs in which particles are initialized with random orientations on a triangular lattice which spans the simulation box. We additionally simulate runs in which the particles are initialized in a hexagonally packed cluster with a) all particles oriented towards the center and b) particles oriented randomly, as well as a run in which the particles are intialized in a persistent worm formation.
We choose these different initial configurations because we see multistabilities, for which the realized state at long times
can depend on the initialization.
We equilibrate the system for $1000\tau_R$ and then collect data for an additional $1000\tau_R$. 

In the following, we vary the strength of the torque interactions, $\mathcal{T}_0$, and the spatial range on which the interactions take place $R$. Throughout, we maintain that $\mathcal{T}_0=\mathcal{T}_A=2\mathcal{T}_{C}$ and that cohesion and alignment interactions take place in the same spatial range. 

\section{Classification of collective behavior}
\label{sec:coll_be}
\begin{figure*}[ht!]
  \centering
  \includegraphics[width=.75\linewidth]{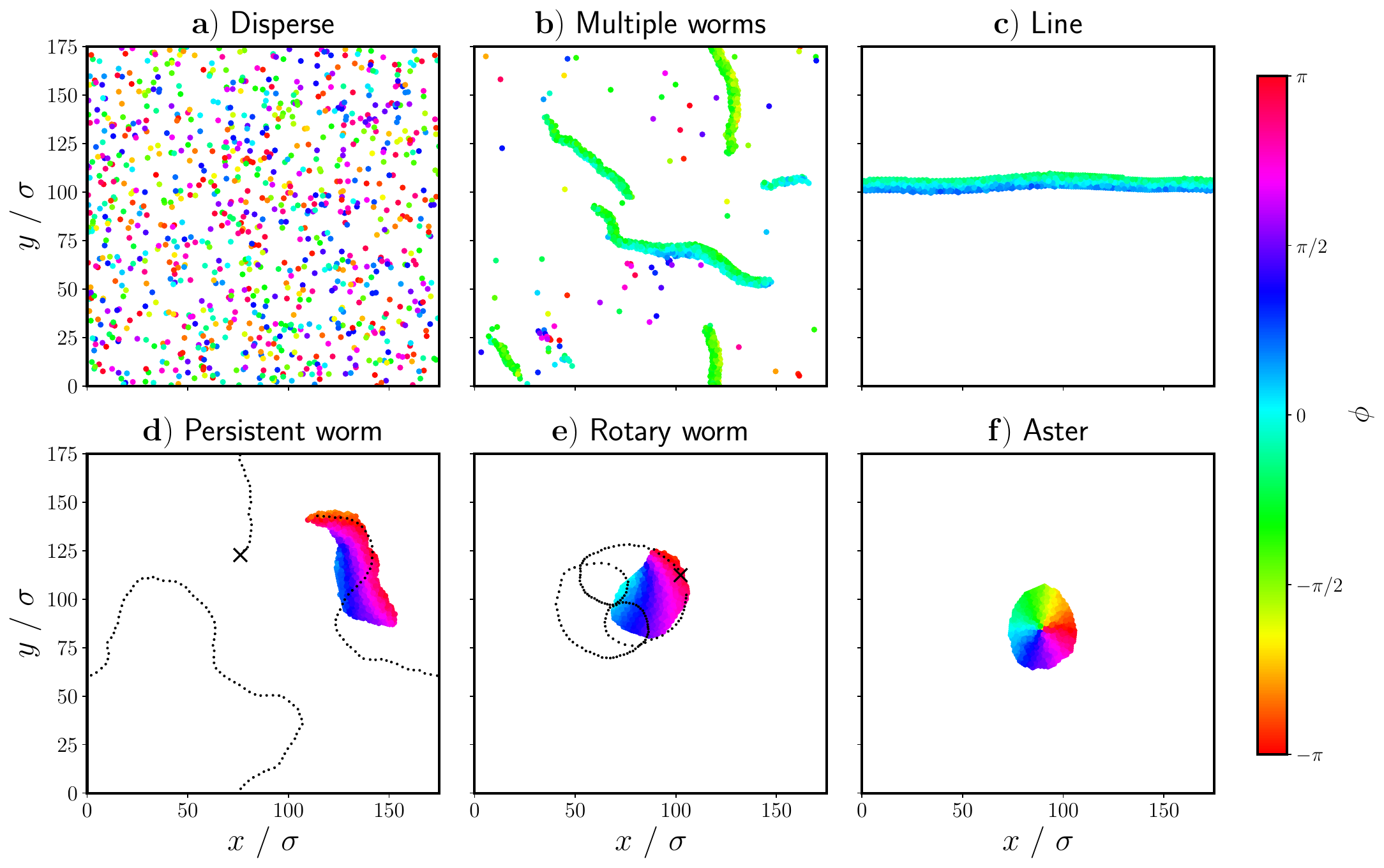}
\caption{Snapshots of the different qualitatively observed states achieved by varying the parameters $R$ and $\mathcal{T}_0$ in Eq.~\eqref{eq:prey-prey_aligngroup}: \textbf{a)} the disperse state, \textbf{b)} the multiple worm state, \textbf{c)} the line state, \textbf{d)} the persistent worm state, \textbf{e)} the rotary worm state, and \textbf{f)} the aster state. Particles are colored based on their orientation, $\phi$. \textbf{d)} and \textbf{e)} additionally show the partial trajectory of a single particle as a dashed line, where the starting position of the particle is marked with a cross.} 
\label{fig:snapshots}
\end{figure*}

By varying the radius of interaction, $R$, and the strength of the interaction torque, $\mathcal{T}_0$, we produce a diverse collection of different behaviors from this model. Qualitatively, we can distinguish six states: a) a disperse state; b) a multiple worm state in which short-lived, worm-like structures form in the presence of a disperse background; c) a line state which spans the simulation box; d) a persistent worm state consisting of a single, persistent worm-like formation; e) a rotary worm state consisting of single or multiple rotating worms which can combine and change direction; and f) an aster state consisting of one or more close-packed clusters in which all particles are oriented towards the center of mass. Snapshots of each of the observed states can be seen in Fig.~\ref{fig:snapshots} and videos can be seen in videos 1-6 of the ESI. In Fig.~\ref{fig:snapshots}, particles are colored based on their orientation. We additionally show the partial trajectory of a single particle as a dashed line in Figs.~\ref{fig:snapshots}d) and e) to illustrate the distinction between the persistent worm state and the rotary worm state. We note that the line state is a finite size effect. In systems with a lower packing fraction, which we did not explore further, we expect it would manifest as a long persistent worm.
This view is supported by our observation, discussed in Sec.~\ref{sec:stat_props}, that for specific parameters the line only transiently forms within the multiple worm state.

A number of these states have previously been observed for the model of Negi \textit{et al.}~\cite{Gompper_similar}. However, the rotary 
worm state is a novel state,  which has, to our knowledge, not previously been observed. Furthermore, our model exhibits multistabilities
between states even when particles are initialized randomly, which we thoroughly
discuss in Sec.~\ref{sec:state_dia}. 
Multistabilities have been recognized in a range of active systems~\cite{Kryuchkov,Kryuchkov2,Kryuchkov3,Gogia}, but to our awareness such multistabilities emerging from random initializations have not been seen or discussed in previous models with cohesive and aligning torques. From the standpoint of nonlinear dynamics, such multistabilities can be viewed as different fixed points and limit cycles of the system.

\subsection{Static properties}
\label{sec:stat_props}

To quantitatively distinguish between these visibly different states, 
we classify the characteristics of the particles' collective behavior. Collective behavior is generally characterized by the fact that individuals move in the same direction and/or that individuals are close in proximity to one another. To describe how well the individuals move in the same direction, we calculate the polar order of all particles for each simulation,
\begin{equation}
\label{eq:polord}
\langle\Psi\rangle_t=\Bigg\langle~\frac{1}{N}\Bigg|\sum^{N}_{i=1}\mathbf{u}_i(t)\Bigg|~\Bigg\rangle_t,
\end{equation}
where $\langle...\rangle_{t}$ represents the time average. 

Furthermore, to describe how close in proximity the individuals are, we calculate the average cluster size $\langle S_c\rangle_t$, which we define as:
\begin{equation}
\label{eq:cluster_size}
    \langle S_c\rangle_t=\Bigg\langle~\frac{1}{N_c(t)}\sum^{N_c(t)}_{c=1}\bigg(\max_{i\in c}~|\mathbf{r}^c_{i}(t)-\mathbf{r}_\mathrm{CoM}^c(t)|\bigg)~\Bigg\rangle_t,
\end{equation}
where $N_c(t)$ is the number of different clusters at time $t$, $\mathbf{r}^c_{i}(t)$ is the position of a particle $i$ within cluster $c$ at time $t$, and $\mathbf{r}_\mathrm{CoM}^c(t)$ is the center of mass of cluster $c$ at time $t$. In order to define $S_c(t)$, we first need to classify clusters in our system. We do this using the data clustering algorithm DBSCAN~\cite{dbscan1,dbscan2} implemented in the machine learning library scikit-learn~\cite{scikit-learn}, where we use a cutoff distance of $1.5\sigma$ and set our minimum number of samples to $1$. For each cluster, we calculate the center of mass, $\mathbf{r}_\mathrm{CoM}^c$, using the method of Ref.~\cite{Bai_Breen_pbc} to account for periodic boundary conditions.

\begin{figure}[ht!]
  \centering
  \includegraphics[width=0.9\linewidth]{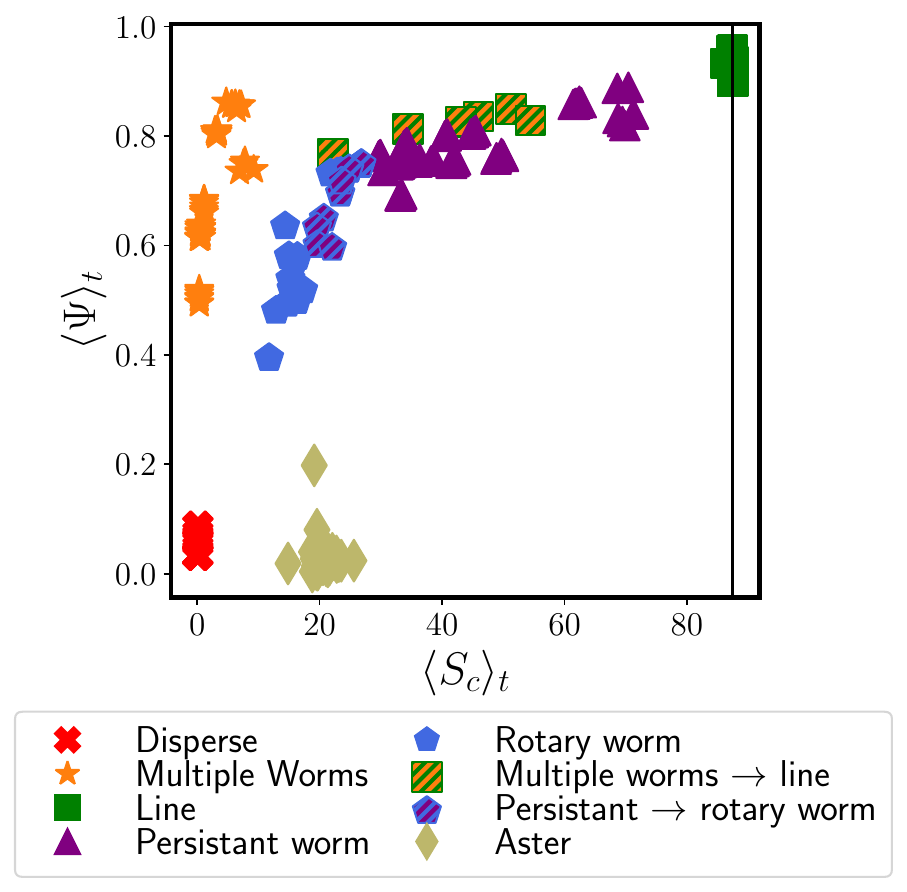}
\caption{Mean polar order $\langle \Psi\rangle_t$ from Eq.~\eqref{eq:polord} plotted vs. the average cluster size $\langle S_c\rangle_t$ 
from Eq.~\eqref{eq:cluster_size}.  Different states and transitions between them
are shown with different colors and symbols. Each symbol refers to one simulation run. The solid vertical line shows half the length of the simulation box.}
\label{fig:group}
\end{figure}

We plot $\langle \Psi\rangle_t$ vs. $\langle S_c\rangle_t$ in Fig.~\ref{fig:group}, from which we are able to distinguish five distinct groupings. We note that in Fig.~\ref{fig:group} we plot each simulation run as a separate point because we observe multistabilities. We identify the grouping with low $\langle \Psi\rangle_t$ and low $\langle S_c\rangle_t$ (shown as red crosses) as the disperse state illustrated in Fig.~\ref{fig:snapshots}a). We can also easily recognize the aster state as the grouping with low polar order but a higher value of $\langle S_c\rangle_t$, shown in Fig.~\ref{fig:group} as khaki diamonds.
The grouping with low cluster size and high polar order can be identified as the multiple worm state (shown as orange stars). 
Despite the fact that multiple worms form independently of one another, they tend to travel in roughly the same direction. 
This results in the relatively high degree of polar order seen in Fig.~\ref{fig:group}. The low cluster size associated with the multiple worm state is due to the disperse background. 
Finally, the grouping with both a high value of $\langle S_c\rangle_t$ and $\langle\Psi\rangle_t$ corresponds to the line state (shown as green squares). 

By defining an average system size $\langle S\rangle_t$ over all particles (see Appendix~\ref{sec:app_syssize}), as opposed to an average cluster size, we additionally find a transition state between the multiple worm and 
line states. In this transition state, multiple worms form a single line which persists for a finite time, but which then breaks up again into multiple worms (see ESI movie 7). The state is depicted as squares with green and orange stripes in Fig.~\ref{fig:group}. The remaining 
data points in Fig.~\ref{fig:group}, which have high values of $\langle\Psi\rangle_t$ and middle ranging values of $\langle S_c\rangle_t$, 
represent a general worm state (depicted by blue circles, purple triangles, and circles with blue and purple stripes). It includes the persistent worms from Fig.~\ref{fig:snapshots}d), which have higher values of $\langle S_c\rangle_t$, as well as the rotary worms from Fig.~\ref{fig:snapshots}e), which have smaller values of $\langle S_c\rangle_t$ and decaying $\langle\Psi\rangle_t$, as well as
transitions between these states.

Based on the static quantities used so far, we have identified five of the six states that we observe qualitatively. However, we are not yet able to distinguish between the persistent and rotary worms. In previous literature~\cite{COUZIN20021,Kryuchkov}, a static milling parameter is defined to characterize rotating states. However, in our case such a parameter is insufficient to identify the rotary worms because their rotational direction
switches over time and the persistent worms also turn (for further explanation, see Appendix~\ref{sec:app_milling}). Therefore, to distinguish between the persistent and rotary worms, as well as to further classify each state, we investigate the dynamic properties of the different states.

\subsection{Dynamic properties}
\label{sec:dyn_props}

\begin{figure} 
  \centering
  \includegraphics[width=1\columnwidth]{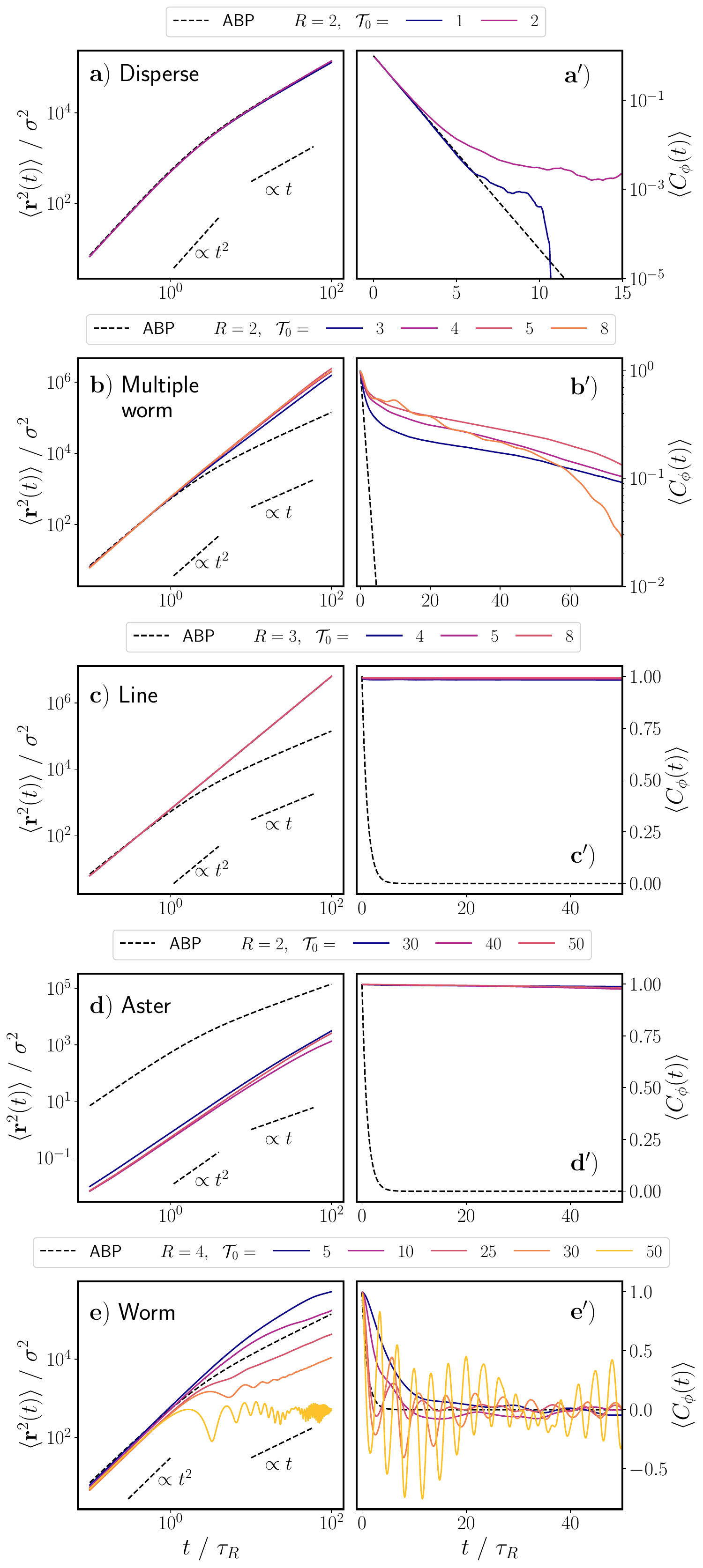}
\caption{Mean squared displacement (Eq.~\eqref{eq:msd}) and orientation correlation functions (Eq.~\eqref{eq:ocf}) for $\mathbf{a)}/\mathbf{a^\prime)}$ the disperse state, $\mathbf{b)}/\mathbf{b^\prime)}$ the multiple worm state, $\mathbf{c)}/\mathbf{c^\prime)}$ the line state, $\mathbf{d)}/\mathbf{d^\prime)}$ the aster state, and $\mathbf{e)}/\mathbf{e^\prime)}$ the general worm state. Note that the x- and y-axis limits are not the same on all graphs. Each row has its own separate legend. Values of $R$ are given in units of $r_c$ and values of $\mathcal{T}_0$ are given in units of $k_\mathrm{B}T$.
}

\label{fig:msd_ocf}
\end{figure}

We investigate the dynamic properties of the different states by calculating the mean squared displacement (MSD):  
\begin{equation}\label{eq:msd}
\langle \mathbf{r}^2(t)\rangle=\frac{1}{N}\sum^{N}_{i=1}\Big\langle(\mathbf{r}_i(t+t_0)-\mathbf{r}_i(t_0))^2\Big\rangle_{t_0},
\end{equation}
and the orientation correlation function (OCF):
\begin{equation}\label{eq:ocf}
\langle C_\phi(t)\rangle=\frac{1}{N}\sum^N_{i=1}\Big\langle\mathbf{u}_i(t+t_0)\cdot\mathbf{u}_i(t_0)\Big\rangle_{t_0},
\end{equation}
of all the active particles in the system. In both Eqs.~\eqref{eq:msd} and~\eqref{eq:ocf}, $N$ is the total number of particles in the system and $\langle...\rangle_{t_0}$ represents the average over time $t_0$ taken after equilibration along the trajectory.
In Fig.~\ref{fig:msd_ocf}, we show exemplary MSD and OCFs for each of the different states identified in Sec.~\ref{sec:stat_props}. Here, we plot the MSD and OCF for a single simulation run, without averaging over further runs, because
of the mentioned multistabilities we observe. In each plot, we additionally show the analytical MSD and OCF of an isolated, standard active Brownian particle (ABP), for which the torque interactions described in Eq.~\eqref{eq:prey-prey_aligngroup} do not exist.

Qualitatively, the disperse state looks very similar to what we would expect for disperse standard ABPs. Indeed, in Fig.~\ref{fig:msd_ocf}a)/a$^\prime$), we see that the calculated MSD and OCF for an active particle in the disperse state match the analytical functions for an isolated ABP very well. The disperse state occurs at low values of $\mathcal{T}_0$ and $R$; therefore, the torque interactions amongst particles are weak and only occur very locally, effectively reducing the behavior of the active particles to that of standard ABPs.

In Fig.~\ref{fig:msd_ocf}b)/b$^\prime$), we see that active particles in the multiple worm state are much more persistent than standard ABPs: the MSD remains ballistic up to times of $100\tau_R$ while for ABPs it crosses over to diffusive motion around $\tau_R$. Correspondingly, the OCF decays much slower than that of a standard ABP. We attribute this persistence to the particles' motion when in a worm formation. Due to alignment interactions, neighboring particles align their orientations parallel to one another, thus moving together in the same direction for a finite amount of time until they break apart.

Similarly to active particles in the multiple worm state, active particles in the line state and the aster state also exhibit a ballistic MSD and a very slowly decaying OCF (see Figs.~\ref{fig:msd_ocf}c)/c$^\prime$) and ~\ref{fig:msd_ocf}d)/d$^\prime$)). In the line state, the particles are even more persistent than in the multiple worm state because the line formation holds throughout the duration of the simulation and has no disperse background. The particles form one group, in which all particles are oriented in approximately the same direction. They then move continuously in that direction throughout the duration of the simulation. In the aster state, the particles also maintain their orientation and remain in the same group throughout the simulation; however, their orientations all point inwards towards the center of mass of the aster. This inwards alignment results in only very slow diffusion of the aster as a whole. Therefore, the MSD of the particles at any given time is significantly lower than that which we would expect for a standard ABP and the OCF decays very slowly. 

In Fig.~\ref{fig:msd_ocf}e)/e$^\prime$), we see that, for worm states which occur at low values of $\mathcal{T}_0$, the MSD and OCF are similar to that which we would expect for a standard ABP, but with a larger persistence. The MSD is ballistic at short times and diffusive at long times, and the OCF decays exponentially for $\mathcal{T}_0=5$ and $10$. The time scale of the transition from ballistic to diffusive behavior (and the exponential decay) is longer for active particles in the worm state as compared with that of a standard ABP. However, as $\mathcal{T}_0$ increases, this time scale becomes shorter. More importantly, we begin to see oscillations in both the MSD and the OCF of the active particles. Eventually, a transition from a ballistic to a diffusive regime in the MSD no longer occurs, but instead we see a transition from ballistic to oscillatory behavior around a plateau value.

The oscillations in the MSD and OCF of the active particles correspond to states in which the worm no longer behaves persistently, as shown in Fig.~\ref{fig:snapshots}d), but instead begins to rotate, as shown in Fig.~\ref{fig:snapshots}e). We use the characteristics of the MSD and OCF to distinguish between the persistent and rotary worm states. We consider worm states classified in Sec.~\ref{sec:stat_props} to be persistent if their MSD transitions from the ballistic to the diffusive regime on a time scale greater than that of a standard ABP. We consider worm states to be rotary if their MSDs transition from ballistic to oscillatory around a plateau value. We additionally define a transition state between persistent and rotary worms (shown by circles with purple and blue stripes in Fig.~\ref{fig:group}; see ESI video 8), for which the MSD still transitions from ballistic to (sub-)diffusive but at a shorter time scale than a standard ABP. The MSD may also exhibit oscillations in this transition state.

\subsection{State diagram}
\label{sec:state_dia}

\begin{figure*}[ht!]
  \centering
  \includegraphics[width=0.95\linewidth]{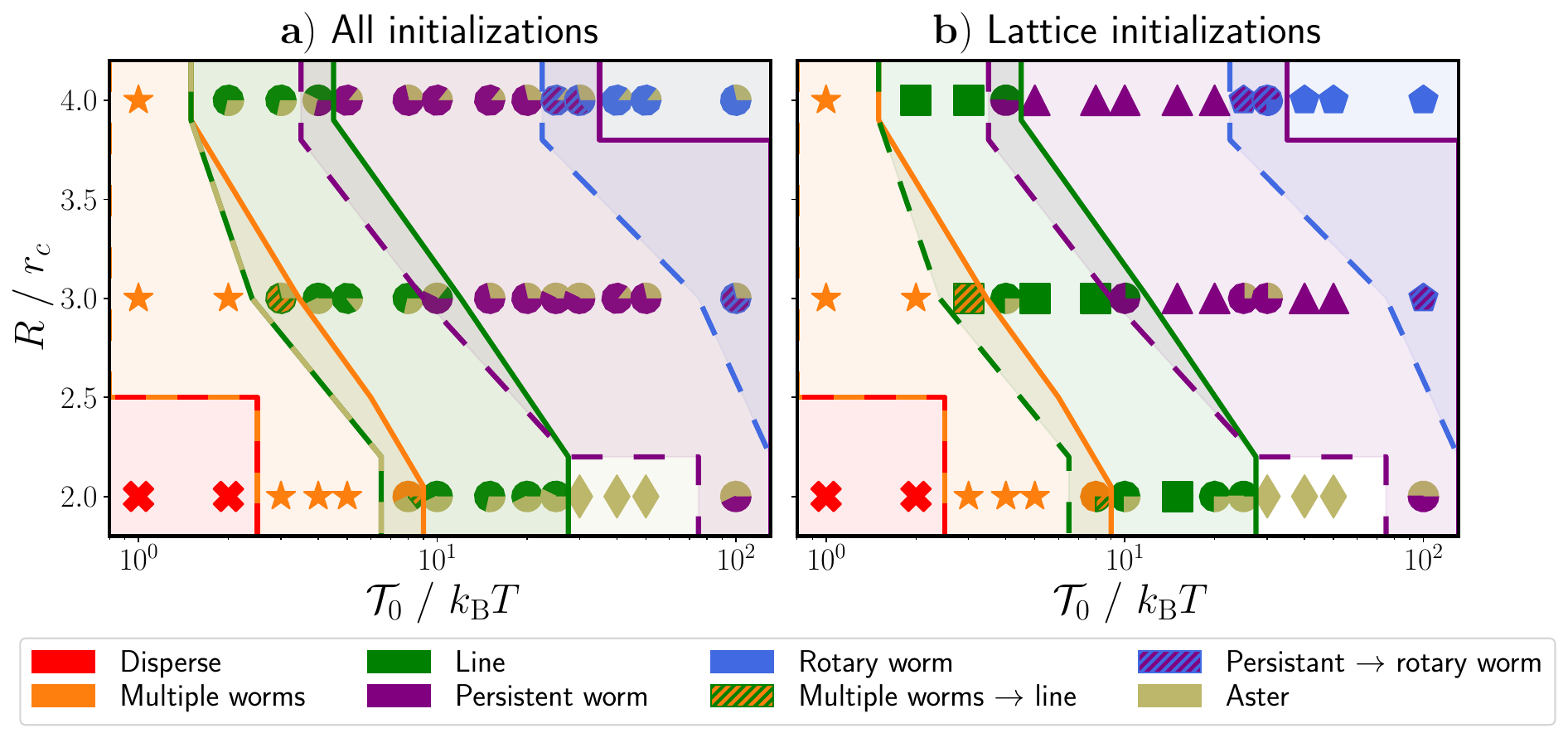}
\caption{State diagram for varying the parameters $R$ and $\mathcal{T}_0$ in Eq.~\eqref{eq:prey-prey_aligngroup}. The different states 
are represented by different colors. In circular markers, the color slices represent the number of occurrences for a given state for a certain 
set of parameters. Markers which are not circular indicate that no multistability exists. 
The number of occurrences is determined $\mathbf{a)}$ using all simulation runs, regardless of initialization conditions, and $\mathbf{b)}$ 
using the four simulation runs in which particles are initialized with random orientations on a lattice.}
\label{fig:state_diagram}
\end{figure*}

Now that we have identified all
possible states of our system within our chosen
range of parameters, we locate them in a state diagram defined by 
the torque strength $\mathcal{T}_0$
and the interaction range $R$. However, as already noted, we observe pronounced multistabilities, for which the realized state depends on 
the initial conditions and stochastics. To more thoroughly understand the impact of initial conditions on the realized state, we initialize the 
particles in various ways: on a lattice with random particle orientations, in an aster, in a hexogonally packed cluster with random particle 
orientations, or in a worm formation.

Figure~\ref{fig:state_diagram} shows the resulting state diagrams, in which the different colored slices of each circular marker represent 
the observed states and the size of each slice indicates the number of realized occurrences for a given state. Markers which are not 
circular indicate that, for that given set of parameters, no multistability exists. Figure~\ref{fig:state_diagram}a) shows the state diagram 
with the number of occurrences when all four initialization types are used, while in Fig.~\ref{fig:state_diagram}b)
the particles are always initialized on a lattice with random orientations. In Fig.~\ref{fig:state_diagram}b), we see that multistabilities emerge even when particles are randomly initialized. Thus, the multistabilities emergent from this model are distinct from the hysteresis observed in the model of Couzin \textit{et al.}~\cite{COUZIN20021,Couzin_2005}.
%
%

Starting from the disperse state in Fig.\ \ref{fig:state_diagram}a) and with increasing $\mathcal{T}_0$, $R$, the system typically shows
a sequence of dispersion, multiple worms, lines, persistent worms, and rotary worms. However, we also see that most states exhibit bistability 
with the aster state. Exceptions occur at very low torques, where either the disperse or multiple worms state exists exclusively and at medium
to high torques for a low interaction radius, where only the aster state occurs.
In the transition regions between states, up to three states can 
coexist for a given set of parameters.

The aster state remains stable, once formed, across many different sets of parameters because this type of formation is maximally compact, thereby satisfying the cohesive tendencies of the particles. It also exhibits local alignment, which satisfies the tendency of particles to align. The only frustration occurs at the center of the aster, because the central particle feels the same torque regardless of its orientation. However, this frustration cannot propagate to the edge of the group. Thus, the outer layers of the aster block the inner, frustrated layers from escaping and the group does not break apart. The stability of the aster state once formed (above a certain interaction-radius-dependent torque strength) is confirmed in Fig.~\ref{fig:app_state_diagram} of Appendix~\ref{sec:aster_stab}, in which we show the state diagram for simulations initialized in an aster formation. We note that, for systems with fewer particles, we expect the aster state to be less stable because the outer layers would be affected by the frustration at the center. Indeed, for systems with the same packing fraction but only 100 particles, aster states only remained stable for sufficiently high torque strengths with an interaction radius of $R=2r_c$ and the exceptional case of a torque strength of $\mathcal{T}_0=100k_\mathrm{B}T$ with an interaction radius of $R=3r_c$ (data not shown).

The aster state only becomes stable once global formations that contain all particles are able to form. Such formations can only occur above a certain 
threshold torque, which depends on the interaction radius. Below this threshold, thermal noise dominates over the interaction torques and any clusters, including asters, that form will be short lived. 

If we compare Fig.~\ref{fig:state_diagram}a) to Fig.~\ref{fig:state_diagram}b), we see that aster states rarely form if particles are initialized with random orientations on a lattice, although it is possible.
From 
such an initialization,
particles can form an aster via two distinct mechanisms: either two worms collide roughly head-on and serve as a nucleation spot for the rest of the particles to gather (see ESI video 9), or a single very long persistent 
worm turns sharply and curls in on itself in a manner similar to an active filament entering a spiral formation (see ESI video 10)~\cite{spiral_activefilament}. 
The second of these mechanisms occurs exclusively 
at medium to high torques and low interaction radii,
for which the aster state is the only stable state. 
In contrast, the first of these mechanisms occurs randomly and causes multistabilities to emerge even when particles are initialized on a lattice with random orientations.

From Fig.~\ref{fig:state_diagram}, we additionally see that different formations generally lie in diagonal bands across the parameter space. 
This means that increasing the interaction radius and increasing the torque strength effectively have the same effect. Furthermore, we expect that for any given interaction radius, the same succession of states can be found. The torque strength at which each state occurs simply changes, becoming lower for high radii of interaction and vice versa.

Now that we have classified the collective behavior of our model into distinct states, we specifically focus on those states which are able 
to both stay in a group and move together in the same direction; namely, the persistent and rotary worm states. We do not further consider 
the line state because it is a finite-size effect.

\section{Properties of worms}
\label{sec:worm_props}
To better understand the collective behavior exhibited by worm states at both a group and an individual level, we analyze the structural properties of the entire worm in addition to the dynamic properties of individuals within the worm.

\subsection{Structural properties}

\begin{figure}
  \centering
  \includegraphics[width=\linewidth]{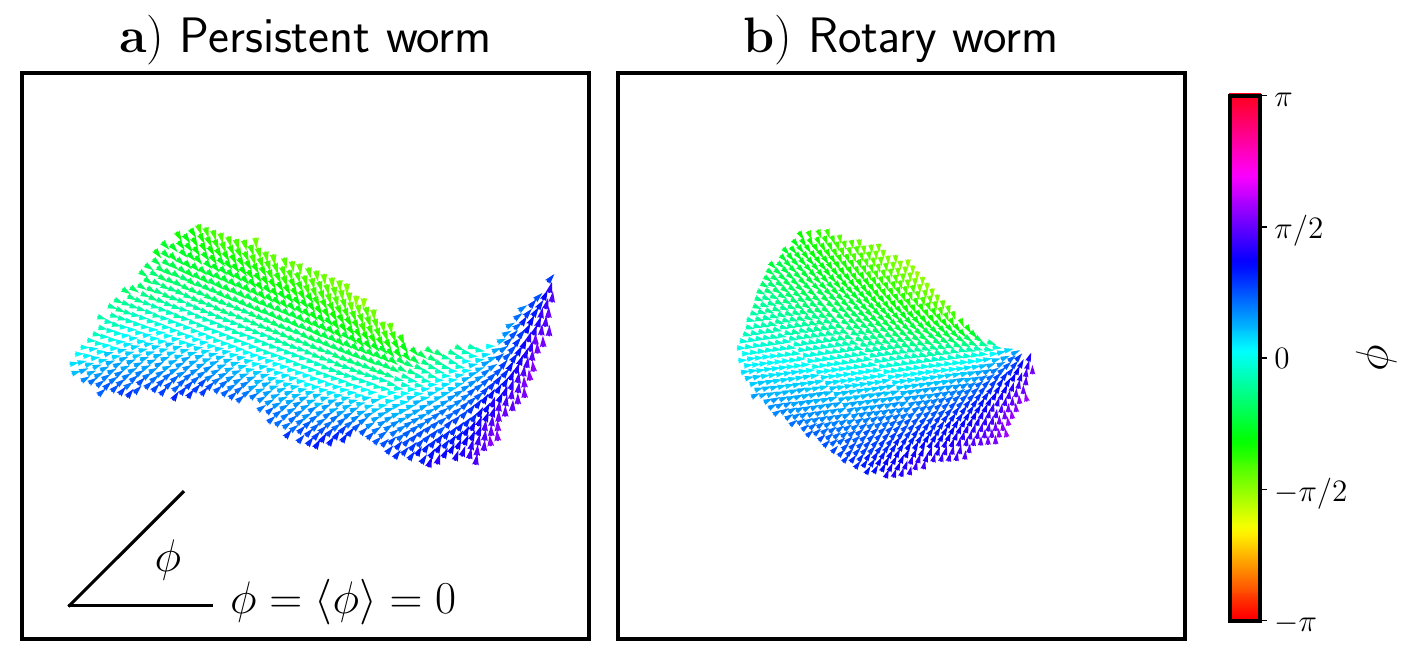}
\caption{Orientations, $\phi$, of particles in \textbf{a)} a persistent worm state and \textbf{b)} a rotary worm state. Particle orientations are shown with respect to the average orientation of the worm $\langle\phi\rangle$ and are indicated by both color and arrowhead direction.}
\label{fig:orientation}
\end{figure}

We begin exploring the structural properties of both persistent and rotary worms by examining their orientation profiles, shown in Fig.~\ref{fig:orientation}. In comparing Figs.~\ref{fig:orientation}a) and b), we see that the orientation profiles of the rotary and persistent worms are very similar. For both types of worms, along the worm's centerline, particles are generally orientation parallel to the worm's average orientation. Here, the cohesive torques from particles on either side cancel out and the particles' tendency to align dominates. 

Particles at the edges of both types of worms, however, point slightly inward towards the centerline. This inward tilted orientation is a result 
of the competition between the alignment and cohesion torques. In Appendix~\ref{sec:app_infsheet}, we perform a stability analysis for the 
idealized case of a particle located at the edge of a uniform sheet. Particles within the uniform sheet have an orientation parallel to the sheet's edge. A schematic of this setup is shown in Fig.~\ref{fig:edge_worm_inf} of Appendix~\ref{sec:app_infsheet}. This analysis shows that the stable orientation for such a particle is approximately
parallel to those of the particles within the sheet, with a slight tilt inward towards the sheet's centerline. This stable orientation only depends
on the ratio between the aligning and cohesive torque strengths.
Qualitatively, this result matches the orientations that we see for particles at the edge of both types of worm states in Fig.~\ref{fig:orientation}.

We also see in comparing Figs.~\ref{fig:orientation}a) and b) that the rotary worm is thicker and more round than the persistent worm. To further explore this difference, we define the asymmetry of each cluster as:
\begin{equation}
    \label{eq:asymm}
    A=\frac{|I_1-I_2|}{I_1+I_2},
\end{equation}
where $I_{1}$ and $I_{2}$ are the principle moments of inertia. A more detailed explanation of this definition can be found in Appendix~\ref{sec:app_symmetry}.

\begin{figure} 
  \centering
  \includegraphics[width=0.75\linewidth]{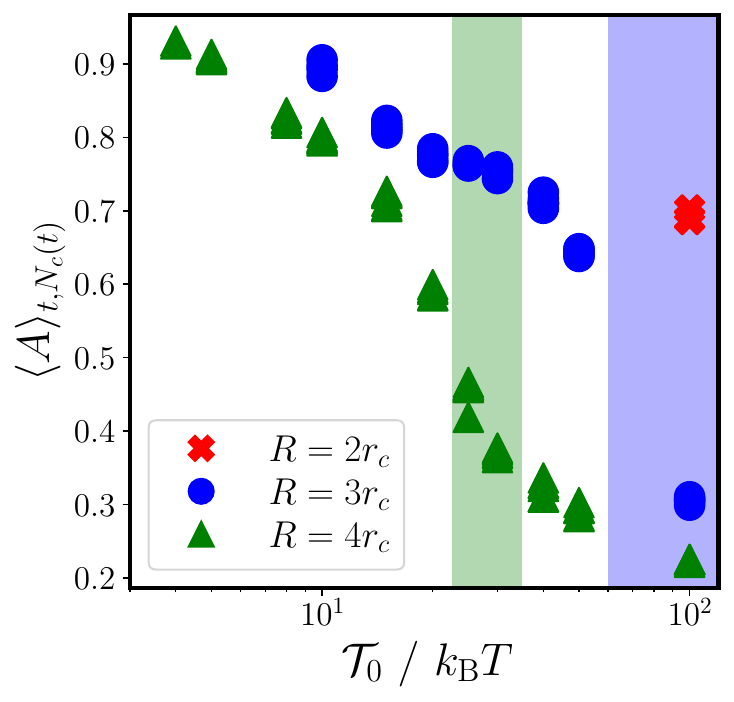}
\caption{Asymmetry (as defined in Eq.~\eqref{eq:asymm}) as a function of $\mathcal{T}_0$ for clusters in the persistent worm state, the 
rotary worm state, and the transition state between the two. Different interaction radii, $R$, are shown as different colors and symbols. 
Vertical shaded regions, colored according to the interaction radius, indicate where the persistent-to-rotary-worm transition occurs.
}
\label{fig:asymmetry}
\end{figure}

Fig.~\ref{fig:asymmetry} shows the time and cluster average of the asymmetry $\langle A\rangle_{t,N_c(t)}$ as a function of the torque strength for all simulation runs that exhibit a worm state, including transition states between persistent and rotary worms. We can clearly see two regimes: one with high asymmetry, which occurs at low torque strengths, and one with low asymmetry, which occurs at high torques. By highlighting the range of torques for which transition states occur in Fig.~\ref{fig:asymmetry}, we see that the high asymmetry regime corresponds to the persistent worm state and the low asymmetry regime corresponds to the rotary worm state. We thus infer that the change in asymmetry of the worm leads to this transition from persistent to rotary worms.

The asymmetry generally decreases with increasing torque strength. As the torque strength increases, the noise in a particle's orientation becomes negligible in comparison, thereby suppressing random deviations in a particle's orientation. These deviations are what allow particles to slip behind or in front of one another to form a more elongated worm. Without such deviations, outer layers of particles simply push on inner layers, without being able to penetrate them. This leads to a more round (symmetric) worm.

For a given torque strength, we also see that the asymmetry is lower for higher interaction radii. To understand this trend, we again consider a particle located at the edge of a sheet; however, this time we consider the sheet to be made up of hexagonally close packed particles oriented parallel to the sheet's edge (see Appendix~\ref{sec:app_hcpsheet}). 
When using this approximation to represent the worm, we find that the angle with which an edge particle points towards the sheet depends on the particle's radius of interaction: as the interaction radius increases, so does the particle's orientation inwards towards the sheet. A larger angle inwards makes it less likely for edge particles to move in front of or behind one another. Instead, edge particles continuously push against their inner neighbors, without being able to move further into the group due to its close packed nature. Thus, a rounder, less elongated worm forms.

\subsection{Dynamics of individuals}

\begin{figure*}
  \centering
  \includegraphics[width=0.75\linewidth]{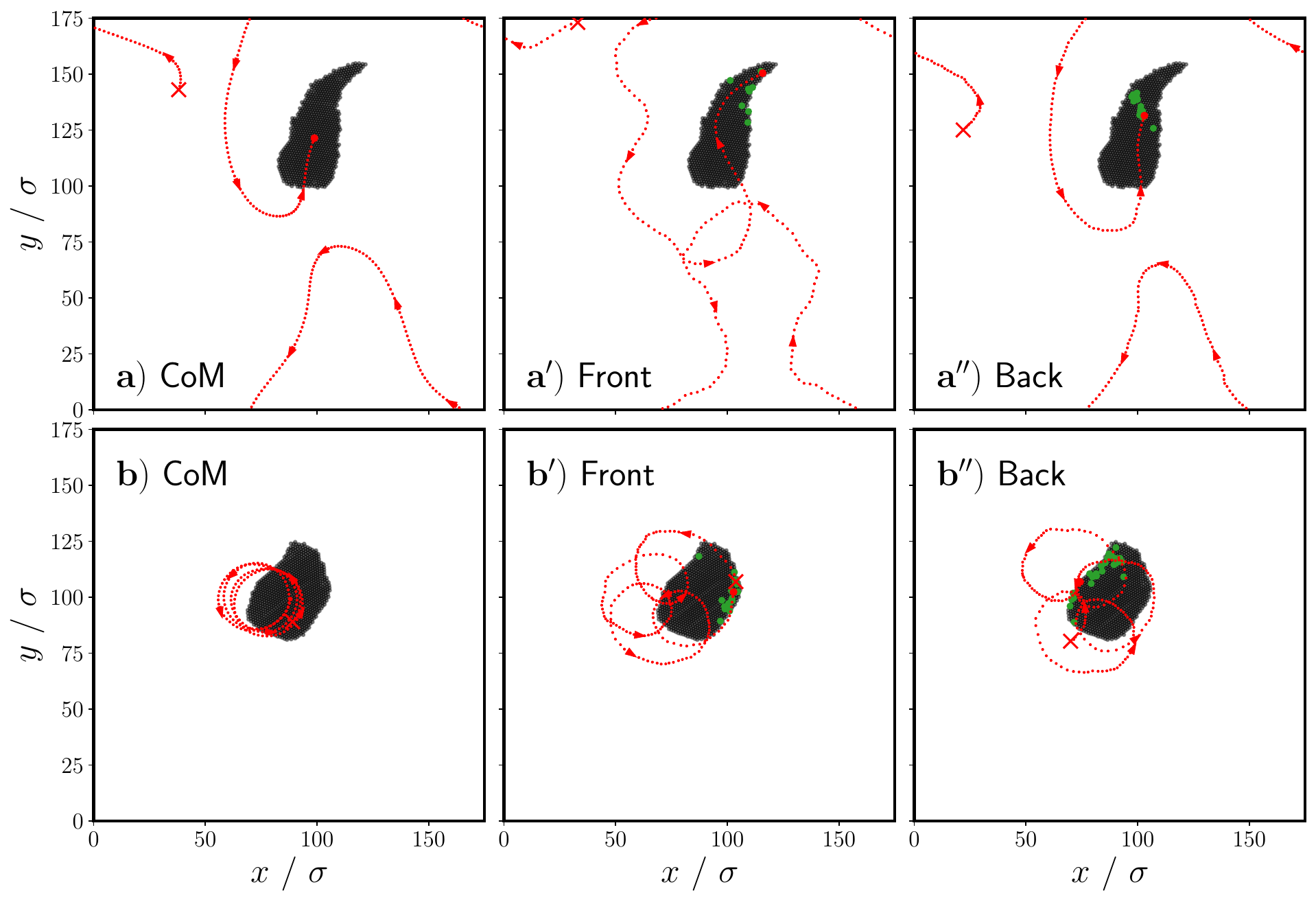}
\caption{Snapshots showing the trajectory of the center of mass of $\mathbf{a)}$ a persistent worm and that of $\mathbf{b)}$ a rotary worm. 
Additionally shown is the trajectory of a single particle $i_f$ at the front of $\mathbf{a^\prime)}$ a persistent worm and $\mathbf{b^\prime)}$ 
a rotary worm, as well as that of a single particle $i_b$ a the back of $\mathbf{a^{\prime\prime})}$ a persistent worm and 
$\mathbf{b^{\prime\prime})}$ a rotary worm. Particles shown in green were, at the beginning of the shown trajectory, within interaction 
distance $R$ of particle $i_f$/$i_b$. Red crosses mark the starting point of each trajectory. Trajectories have a duration of $20\tau_R$.}
\label{fig:worm_traj}
\end{figure*}

A crucial feature of collective behavior in animal groups is that neighbors are dynamic and change as individuals move~\cite{change_neighbors}. To assess whether our collective behavior model has dynamically changing neighbors, we explore the dynamics of individuals and their neighbors within the group. In Fig.~\ref{fig:worm_traj}, for both persistent and rotary worms, we plot the trajectory of an individual particle $i_f$ which begins at the front of the group (Figs.~\ref{fig:worm_traj}a$^\prime)$ and b$^\prime)$) and an individual particle $i_b$ which begins at the back of the group (Figs.~\ref{fig:worm_traj}a$^{\prime\prime})$ and b$^{\prime\prime})$). We additionally show, in green, particles which were, at the beginning of the shown trajectory, within interaction distance $R$ of particle $i_f$ or $i_b$. For comparison, we show the center of mass (CoM) trajectories for both types of worms in Figs.~\ref{fig:worm_traj}a) and b).

We immediately see that neighbors change dynamically over time: neighboring particles at the beginning of the trajectory are no longer neighbors at the end. Furthermore, we see that particle $i_f$, which begins as a ``leader" at the front of the group, is no longer in the lead. This phenomenon of changing leadership is also seen in the collective behavior of animals~\cite{bird_leaders}.

Comparing the different trajectories shown in Figs.~\ref{fig:worm_traj}a), a$^\prime$), and a$^{\prime\prime}$), we see that the entire trajectory of particle $i_b$ is very similar to that of the CoM. However, the trajectory of particle $i_f$ differs in that it is much less straight and has many more curves along it. We perform a stability analysis similar to that done for particles at the edge of worms, but now we model the worm as a uniform sheet shaped like an isosceles triangle and oriented along its axis of symmetry (see schematic in Fig.~\ref{fig:front_worm} of Appendix~\ref{sec:app_frontworms}). We investigate the behavior of a particle at the foremost tip along the axis of symmetry as a means to understand the behavior of particles at the front of persistent worms.
From this stability analysis, we find that the stable orientation for such a particle is parallel to the orientation of the particles in the sheet. The orientation will therefore return to this one after any small deviation due to noise, leading to many small curvatures along the trajectory as can be seen in Fig.~\ref{fig:worm_traj}a$^\prime$).

For the rotary worm, trajectories for both particles $i_f$ and $i_b$ differ significantly from that of the CoM. Whereas the CoM shows a roughly circular trajectory, particles $i_f$ and $i_b$ exhibit looping trajectories which resemble that expected for a particle precessing around a point. These trajectories indicate that individual particles are orbiting the center of mass, which simultaneously orbits a central point. 

We conclude this section with the comment that, although particles in worm states ostensibly move together and can be considered as a single body, in fact within this body individuals follow unique trajectories. These distinct dynamics of individuals are in contrast to the railway motion seen in the model of Negi \textit{et al.} for individuals in worm-like swarms~\cite{Gompper_similar}.

\section{Conclusions and outlook}
\label{sec:conc_out}

We have explored the emergent structural and dynamic properties of particles that interact via alignment and non-reciprocal cohesive 
torques. Depending on their radius of interaction and the strength of their interaction torques, particles exhibit different types of collective behavior, ranging from multiple, short-lived worms in a disperse background to single, 
long-lived worms, which can exhibit either rotary or persistent dynamics to closely packed asters. By analyzing both static and dynamic properties of the observed collective behaviors, we classified six distinct possible states for this model. The realized state for a given set of parameters is, in many 
cases, multistable and depends on stochastics and initial conditions.

The persistent and rotary worm states  that emerge from this model are particularly emblematic of collective behavior because, in these states, particles both move in the same direction and stay together in a single group. Furthermore, these states exhibit continuously changing neighbors and leadership, both of which are observed in animal groups such as flocks of birds~\cite{bird_leaders,change_neighbors}. 
Qualitatively, similar worm-like structures can be observed in many examples of collective behavior seen in nature~\cite{birdworm1,slug_cells,slugs_cells2}. In future work, it would be interesting to quantitatively compare the dynamics and structures exhibited by this model to those exhibited by specific systems in nature. In addition, this model could also serve as a template for engineering collective behavior in microrobotic systems~\cite{microrobots1,microrobots2}, which is desirable for applications in fields such as biomedicine. Our model
is particularly conducive to this application because the multistable nature
of many states should allow for reconfigurability.

Worm-like structures similar to our persistent worm state have previously been observed for the collective behavior models devised by Couzin \textit{et al.}~\cite{COUZIN20021,Couzin_2005} and Negi \textit{et al.}~\cite{Gompper_similar}. However, to our awareness, a state analogous to our rotary worm state has not previously been observed. Additionally, Negi \textit{et al.} analyzed the behavior of their persistent worm-like structure through the lens of railway motion performed by active polymers~\cite{spiral_activefilament}. Such a lens is insufficient to describe the worm state exhibited by our model, due to dynamics within the worm itself. A further contrast to the models of Couzin \textit{et al.} and Negi \textit{et al.} is that we observe multistabilities in many of our states even when the particles are initialized randomly. Such multistabilites were not observed in these previous models.

In this work, we have focused on varying the strength of the torque interaction and the radius of this interaction, which applies to both the 
alignment and cohesive torques. However, we have always kept the ratio between the alignment and cohesion torques constant. 
In future work, it would be interesting to vary this ratio as well as to use different spatial ranges for the alignment and cohesive torques.
This would enable further exploration of the analogy between the observed collective dynamics in the model and similar emergent behavior in nature. It would also broaden the scope of interactions which could be used to engineer swarms of microrobots for applications in biomedical, among other, fields.

\backmatter

\bmhead{Supplementary information}

Below is the link to the electronic supplementary material.

\bmhead{Acknowledgements}
We thank Till Welker, Josua Grawitter, Swarnajit Chatterjee, and Fernando Peruani for helpful discussions. We acknowledge financial support from TU Berlin and the German Research Foundation (DFG) as a part of research grant number 462445093. We acknowledge support by the Open Access Publication Fund of TU Berlin.

\section*{Declarations}
\subsection*{Funding}
Open Access funding enabled and organized by Projekt DEAL.
\subsection*{Competing interests}
There are no conflicts of interest to declare.
\subsection*{Data availability}
The datasets generated and/or analyzed during the current study are available from the corresponding author on reasonable request.

\begin{appendices}

\section{Average system size}
\label{sec:app_syssize}
Analogously to the average cluster size (Eq.~\eqref{eq:cluster_size}), we define the average system size as:
\begin{equation}
\label{eq:sys_size}
    \langle S\rangle_t=\Big\langle~\max_{i}~|\mathbf{r}_{i}(t)-\mathbf{r}_\mathrm{CoM}(t)|~\Big\rangle_t,
\end{equation}
where $\mathbf{r}_\mathrm{CoM}(t)$ is the center of mass of all the particles in the system at a time $t$ and $\mathbf{r}_{i}(t)$ is the position of particle $i$ at time $t$. We note that $\langle S\rangle_t$ is a measure of the distance among all particles, whereas the previously discussed
$\langle S_c\rangle_t$ is a measure of the distance only among particles within the same cluster. This means that states with disperse particles have a low value of $\langle S_c\rangle_t$, because the average cluster size is small, but a high value of $\langle S\rangle_t$, because the particles are spread out.

Plotting $\langle S\rangle_t$ vs. $\langle \Psi\rangle_t$ reveals distinct groupings for the disperse (red crosses) and aster states (khaki diamonds). We note that there is an outlier for the aster state grouping, for which the value of $\langle S\rangle_t$ is significantly higher than for other aster states. This outlier, which we do not see in the value of $\langle S_c\rangle_t$, occurs because in this particular simulation two distinct asters formed and did not combine for the duration of the simulation. For the remaining three
states, which we were able to distinguish in Fig.~\ref{fig:group}, it is not possible to establish a clear threshold to separate them from each other.

However, plotting $\langle S\rangle_t$ vs. $\langle \Psi\rangle_t$ does reveal a transition state between the multiple worm and worm states, depicted in Fig.~\ref{fig:group2} as squares with green and orange stripes. Although these simulations fall into the general worm state based on values of $\langle S_c\rangle_t$ (see Fig.~\ref{fig:group}), in Fig.~\ref{fig:group2} we see 
that their values of $\langle S\rangle_t$ are more similar to those of the multiple worm state. The reason for the discrepancy is that, in this transition state, which we already described in Sec.~\ \ref{sec:stat_props},
particles form a line which spans the simulation box. However, this line becomes unstable and eventually breaks apart into multiple worms. Therefore, over time, the value $S_c(t)$ oscillates between a value which corresponds to a multiple worm state and a value which corresponds to a line state, resulting in an average value similar to that which we would expect for the general worm state.

\begin{figure}
  \centering
  \includegraphics[width=0.8\linewidth]{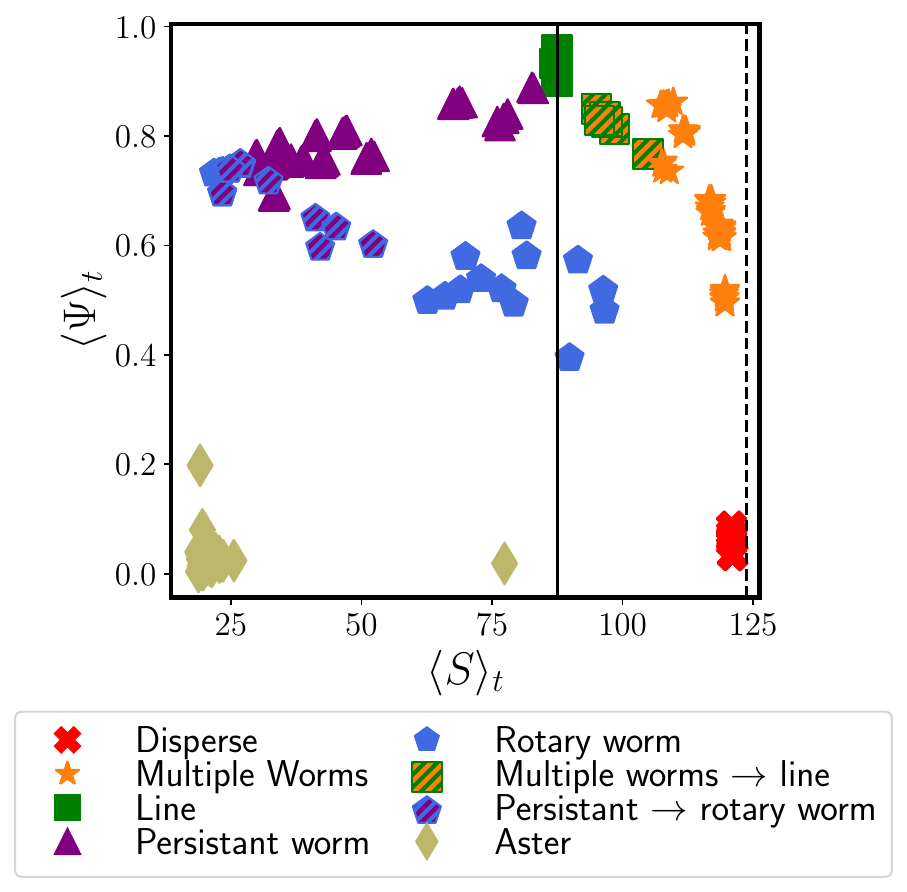}
\caption{Average system size ($\langle S\rangle_t$,
Eq.~\eqref{eq:sys_size})
vs. polar order ($\langle \Psi\rangle_t$, Eq.~\eqref{eq:polord}). Different formation classifications are shown with different colors and symbols. The solid vertical line shows half the length of the simulation box and the dashed vertical line shows the maximum distance two particles can be from one another.}
\label{fig:group2}
\end{figure}

\section{Milling parameter}
\label{sec:app_milling}

To introduce a milling parameter, we first define the time dependent quantity:
\begin{equation}
\label{eq:milling}
    M(t,N_c(t))=\frac{1}{n_c}\sum^{n_c}_{\substack{i=1,\\i\in c}}\Bigg(\frac{\mathbf{r}^c_i(t)-\mathbf{r}^c_\mathrm{CoM}(t)}{|\mathbf{r}^c_i(t)-\mathbf{r}^c_\mathrm{CoM}(t)|}\times\mathbf{u}_i(t)\Bigg),
\end{equation}
where $N_c(t)$ is the number of different clusters at time $t$, $n^c$ is the number of particles in cluster $c$, $\mathbf{r}^c_{i}(t)$ is the position of a particle $i$ which is a constituent of cluster $c$ at time $t$, and $\mathbf{r}_\mathrm{CoM}^c(t)$ is the center of mass of cluster $c$ at time $t$. We use the same clusters defined in Sec.~\ref{sec:stat_props}. We consider all clusters with $n^c>1$.

We now define a milling parameter $|\langle M\rangle_{t,N_c(t)}|$ by taking the absolute value of the time and cluster average of this quantity. This milling parameter is very similar to those defined in Refs.~\cite{COUZIN20021,Kryuchkov}. We note that, in these references, the milling states identified generally have 
the shape of a ring, in which case no particles are located at the center of mass, in contrast to our rotary worm states.

\begin{figure}
  \centering
  \includegraphics[width=\linewidth]{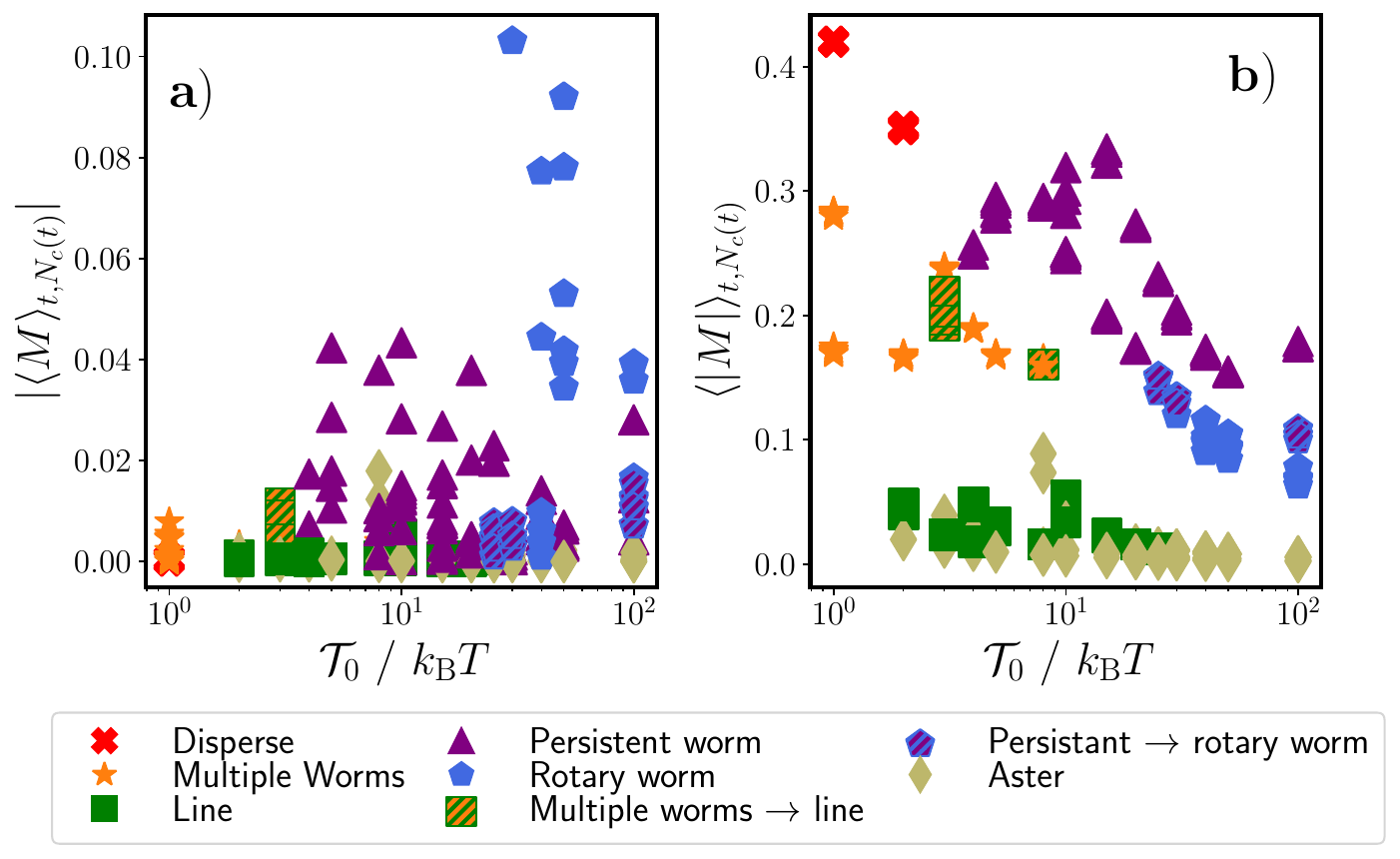}
\caption{Two different milling parameters, based on Eq.~\eqref{eq:milling}, as a function of $\mathcal{T}_0$: \textbf{a)} $|\langle M\rangle_{t,N_c(t)}|$ and \textbf{b) $\langle |M|\rangle_{t,N_c(t)}$}. Each different state is represented by a different color and marker.}
\label{fig:millparam_def}
\end{figure}

We plot $|\langle M\rangle_{t,N_c(t)}|$
as a function of $\mathcal{T}_0$ in Fig.~\ref{fig:millparam_def}a). From this, we see that although the value of $|\langle M\rangle_{t,N_c(t)}|$ is high for rotary worms in some cases, in other cases it is relatively low. In Fig.~\ref{fig:snapshots}e) the trajectory of a particle in a rotary worm resembles that which we expect for a point on a precessing body. Therefore, the 
instantaneous value of $M(t,N_c(t))$ for any given cluster is high. However, the rotating worm can spontaneously change direction, meaning that due to the time averaging $|\langle M\rangle_{t,N_c(t)}|$ is not necessarily large. Furthermore, if there are multiple rotary worms, they need not rotate in the same direction. This also contributes to a reduced value of $|\langle M\rangle_{t,N_c(t)}|$ in certain cases.

This observation suggests that we should take the absolute value of Eq.~\eqref{eq:milling} before averaging, thus defining the milling parameter $\langle |M|\rangle_{t,N_c(t)}$, which is shown as a function of $\mathcal{T}_0$ in Fig.~\ref{fig:millparam_def}b). In Fig.~\ref{fig:millparam_def}b), we see that $\langle |M|\rangle_{t,N_c(t)}$ is consistently higher for persistent worms in comparison with rotary worms. As Fig.~\ref{fig:snapshots}d) shows, particles in a persistent worm can at times have a very curved trajectory,
where the absolute value of $M(t,N_c(t))$ 
becomes large. As a result, the values of $\langle |M|\rangle_{t,N_c(t)}$ are large also for persistent worms. 


Figure~\ref{fig:millparam_def}b) also shows that disperse and multiple worm states have high values of $\langle |M|\rangle_{t,N_c(t)}$. This high value results from very small and short lived clusters, in which $\mathbf{r}^c_i(t)-\mathbf{r}^c_\mathrm{CoM}(t)$ and $\mathbf{u}_i(t)$ are generally approximately perpendicular to one another. To eliminate these cases, we could limit the clusters which we use for our definition of the milling parameter to those with higher $n_c$. However, since it is already clear that this parameter is insufficient to distinguish rotary from persistent worms, we do not further refine our definition of $\langle |M|\rangle_{t,N_c(t)}$.

\section{Stability of asters}
\label{sec:aster_stab}
\begin{figure}
  \centering
  \includegraphics[width=0.95\linewidth]{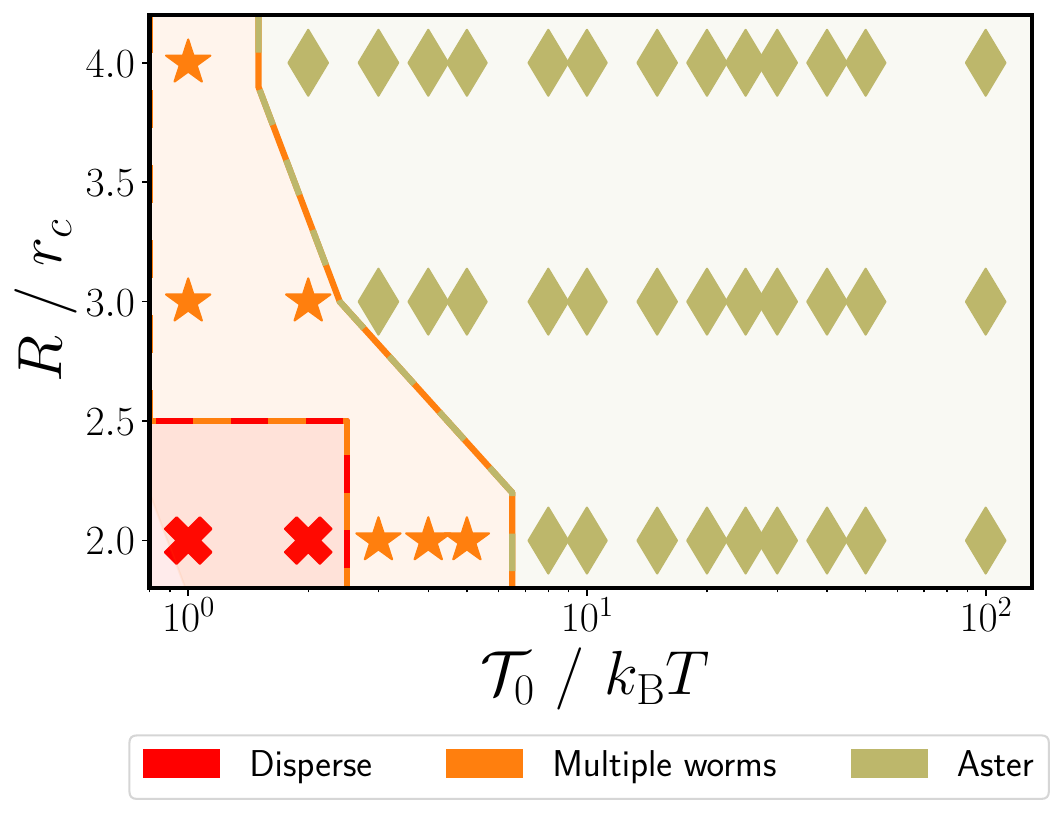}
\caption{State diagram for varying the parameters $R$ and $\mathcal{T}_0$ in Eq.~\eqref{eq:prey-prey_aligngroup} when particles are initialized in an aster state. The different states are represented by symbols with different colors and shapes.}
\label{fig:app_state_diagram}
\end{figure}

To further investigate the stability of the aster state and, more generally, the multistabilities in our system, we run simulations for which particles are initialized in an aster state, i.e. in a hexagonally close packed group in which all particles are oriented towards the center of mass. Figure~\ref{fig:app_state_diagram} is a state diagram of the realized states for this initialization. We see that, above a certain interaction-radius-dependent torque strength, the aster state remains stable. Figure~\ref{fig:app_state_diagram} supports the statement in Sec.~\ref{sec:state_dia} that the aster state becomes very stable once interaction torques dominate over orientational noise.

\section{Stability analysis for the edge of a worm}
\label{sec:app_edgeworms}

\subsection{Uniform sheet}
\label{sec:app_infsheet}
To explore the behavior of a particle at the edge of a worm, we model the worm as a uniformly dense sheet (density $\rho$), which extends infinitely in the x-direction and infinitely above the particle of interest, which sits at its lower edge. All particles within the uniform sheet are oriented along the $x$-axis, such that $\phi_j=0$ for all $j$. The particle of interest, particle $i$, is situated at the edge of the worm, as shown in Fig.~\ref{fig:edge_worm_inf}, and has orientation $\mathbf{u}_i=(\cos\phi_i,\sin\phi_i)$. We would like to determine the orientation of this particle when it interacts with the sheet via the torques given in Eq.~\eqref{eq:prey-prey_aligngroup}.
\begin{figure}
  \centering
  \includegraphics[width=0.85\linewidth]{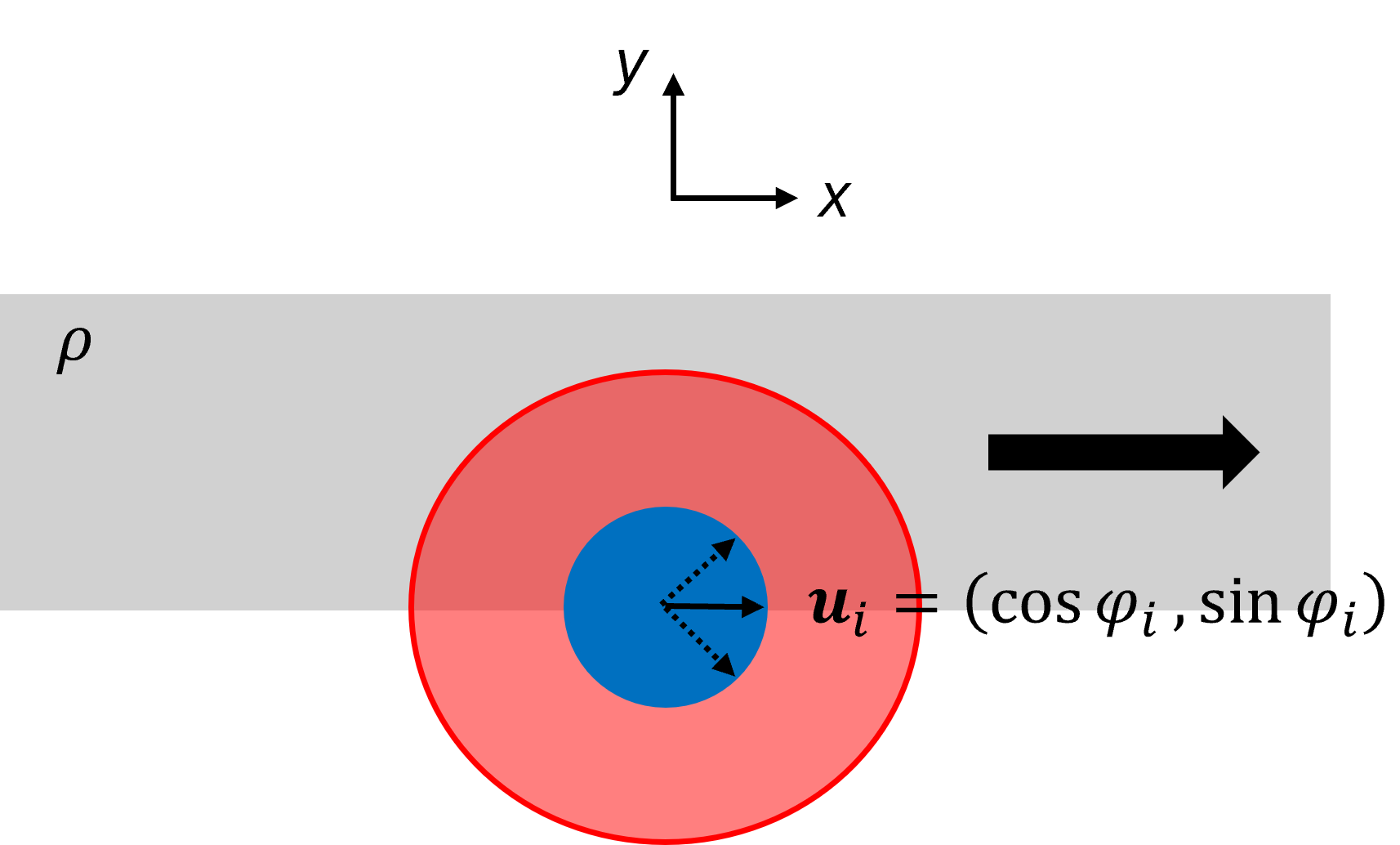}  
\caption{Schematic of a particle at the edge of a worm, where the worm is approximated by an infinite sheet of uniform density oriented along the x-axis. The particle of interest is shown in blue and the radius of interaction is represented by a red concentric circle around the particle.}
\label{fig:edge_worm_inf}
\end{figure}

Using Eq.~\eqref{eq:prey-predy_align} and the fact that $\phi_j=0$, we can write down the alignment torque on particle $i$ as:
\begin{align}
\label{eq:align_sheet}
\mathcal{T}^A_{ij}&=-\mathcal{T}_A\int^R_0\mathrm{d}r'\int^\pi_0\mathrm{d}\phi'~\rho r' \sin{(\phi_i)}\nonumber\\
&=-\frac{\rho\pi R^2}{2}\mathcal{T}_A\sin{(\phi_i)}.
\end{align}
In Eq.~\eqref{eq:align_sheet}, the variables of integration, $r'$ and $\phi'$, represent the radial distance from particle $i$ and the angle from the x-axis respectively. Thus, the double integral in each equation goes over the area within the radius of interaction of particle $i$. The integral over $\phi'$ only goes from $0$ to $\pi$, in spite of the fact that no vision cone is used, because no particles are present from angles $\pi$ to $2\pi$.

Particle $i$ additionally experiences a cohesion torque (using Eq.~\eqref{eq:prey-prey_group}):
\begin{align}
\label{eq:cohesion_sheet}
\mathcal{T}^C_{ij}&=\mathcal{T}_C\int^R_0\mathrm{d}r'\int^\pi_0\mathrm{d}\phi'~\rho r' \Big(\cos{(\phi_i)}\sin{(\phi')} \nonumber\\
&\quad-\sin{(\phi_i)}\cos{(\phi')}\Big)\nonumber\\
&=\rho R^2\mathcal{T}_C\cos{(\phi_i)}.
\end{align}
The variables of integration in Eq.~\eqref{eq:cohesion_sheet} are the same as those in Eq.~\eqref{eq:align_sheet}.

We now let $\mathcal{T}_C=\mathcal{T}_0$. To correspond to our simulation results, we assume that $\mathcal{T}_A/\mathcal{T}_{C}=2$. Thus, we can write the total torque, $\mathcal{T}_{ij}$, on the particle as:
\begin{equation}
\label{eq:}
\mathcal{T}_{ij}=\rho R^2\mathcal{T}_0\Big(\cos{(\phi_i)}-\pi\sin{(\phi_i)}\Big)
\end{equation}
The particle feels no torque when $\mathcal{T}_{ij}=0$, which occurs when the particle has an orientation $\phi_0=\pi n + \arctan{(1/\pi)}$, where $n$ is either $0$ or $1$ (theoretically it can be any integer, but physically it is limited to these two values).

We now determine the stability of $\phi_0$ by looking at the behavior of $\mathcal{T}_{ij}$ near $\phi_0$,
\begin{equation}
\label{eq:stability_edge}
\left.\frac{\partial \mathcal{T}_{ij}}{\partial \phi_i}\right\vert_{\phi_i=\phi_0}=-\rho R^2\mathcal{T}_0\Big(\sin{(\phi_0)}+\pi\cos{(\phi_0)}\Big).
\end{equation}
Evaluating Eq.~\eqref{eq:stability_edge} for our two values of $n$, we find that for $\phi_0=\arctan{(1/\pi)}$, $\left.\frac{\partial \mathcal{T}_{ij}}{\partial \phi_i}\right\vert_{\phi_i=\phi_0}<0$; therefore, this solution is stable. For $\phi_0=\pi+\arctan{(1/\pi)}$, $\left.\frac{\partial \mathcal{T}_{ij}}{\partial \phi_i}\right\vert_{\phi_i=\phi_0}>0$; therefore, this solution is unstable.

The stable solution $\phi_0=\arctan{(1/\pi)}$ means that the stable orientation for this particle is pointed along the x-axis with a slight upward tilt towards the sheet.


\subsection{Hexagonally close packed sheet}
\label{sec:app_hcpsheet}

In the previous section, we approximate the worm as an infinite sheet of uniform density to explore the orientational behavior of a particle at the edge of a worm. 
We now build upon this approximation by treating the worm as 
an infinite sheet of hexagonally packed particles
(see schematic in Fig.~\ref{fig:edge_worm_hcp}). We place our particle of interest at the origin, such that the other particles are located at superpositions of the vectors $\mathbf{a}_1=r_c\hat{\mathbf{x}}$ and $\mathbf{a}_2=r_c(\frac{1}{2}\hat{\mathbf{x}}+\frac{\sqrt{3}}{2}\hat{\mathbf{y}})$.  
\begin{figure}
  \centering
  \includegraphics[width=0.85\linewidth]{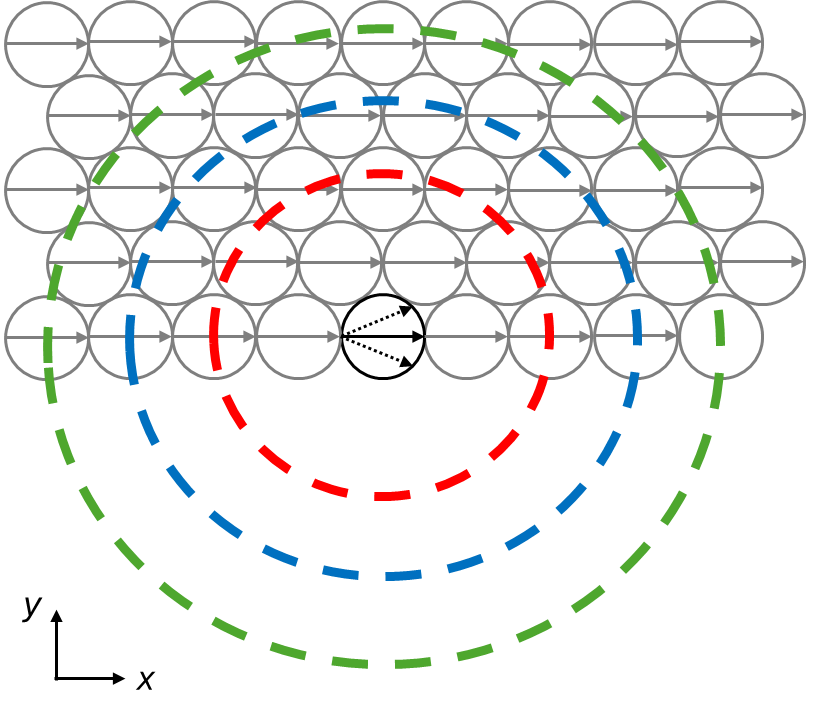}  
\caption{Schematic of a particle at the edge of a worm, where the worm is approximated by an infinite sheet of hexagonally close packed particles oriented along the x-axis. Concentric circles indicate different radii of interaction (red: $R=2r_c$, blue: $R=3r_c$, green: $R=4r_c$) of the particle of interest.}
\label{fig:edge_worm_hcp}
\end{figure}

We then numerically calculate the torque on our particle of interest for different radii of interaction based on Eq.~\eqref{eq:prey-prey_aligngroup}. Angles for which the torque is zero are shown in Fig.~\ref{fig:angles_hcp}a). By comparing these angles, we see that as the radius of interaction $R$ increases, so does the angle of edge particles towards the sheet. As $R\to\infty$, the orientation approaches that determined analytically in Appendix~\ref{sec:app_infsheet} for a particle at the edge of a uniform sheet.

If we do the same analysis for the first inner layer of particles, using the orientations determined previously for the outermost layer, we find that the difference in angle between different interaction radii becomes even more pronounced. The results of this analysis can be seen in Fig.~\ref{fig:angles_hcp}b).

\begin{figure}
  \centering
  \includegraphics[width=\linewidth]{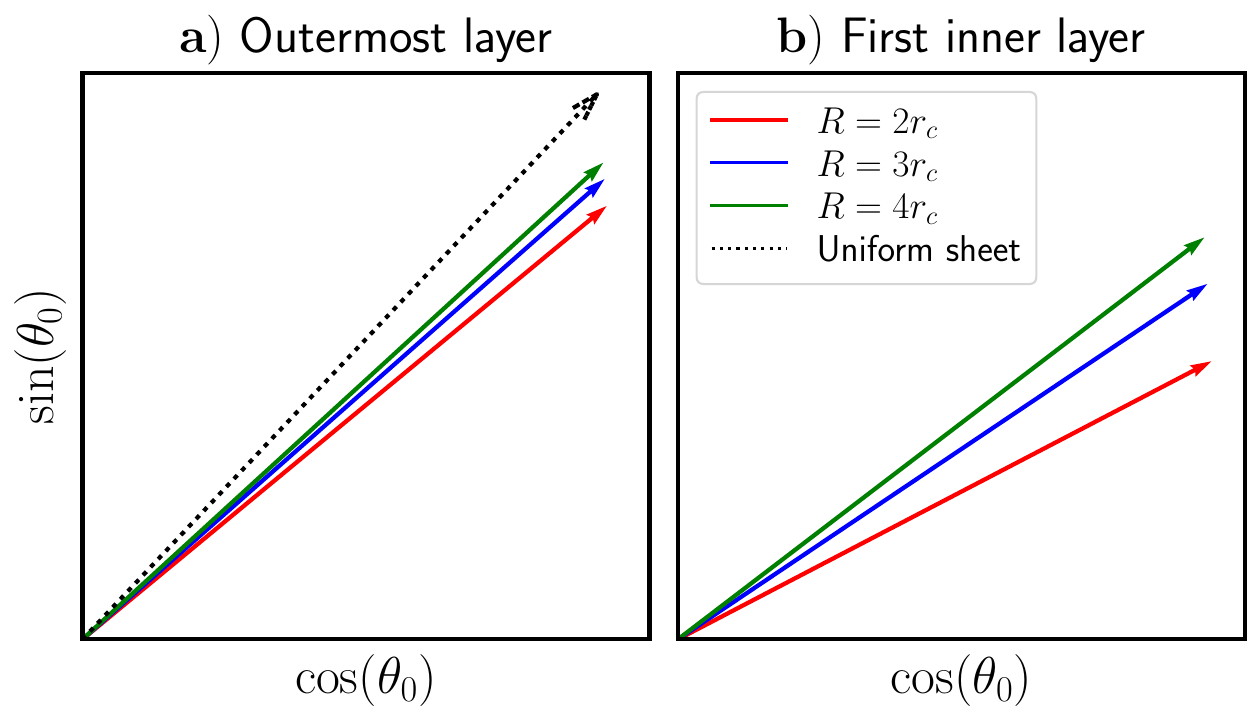}  
\caption{Orientations of particles \textbf{a)} in the outermost layer and \textbf{b)} in the first inner layer of a hexagonally close packed sheet (see schematic in Fig.~\ref{fig:edge_worm_hcp}). The dotted line in \textbf{a)} shows the orientation for a particle in the outermost layer as determined in Appendix~\ref{sec:app_infsheet}.}
\label{fig:angles_hcp}
\end{figure}

\section{Symmetry of clusters}
\label{sec:app_symmetry}
To determine the symmetry of clusters, we begin by calculating the moment of inertia tensor for each cluster:

\begin{equation}
    \mathbf{I}=
\begin{bmatrix}
I_{xx} & I_{xy}\\
I_{yx} & I_{yy}
\end{bmatrix}
\end{equation}
with $I_{xx}=m\sum_{i=1}^{N_c}y_i^2$, $I_{yy}=m\sum_{i=1}^{N_c}x_i^2$, and $I_{xy}=I_{yx}=-m\sum_{i=1}^{N_c}x_iy_i$. Here, $x_i$ and $y_i$ are calculated with respect to the center of mass of the cluster. We calculate the asymmetry for all clusters defined in Sec.~\ref{sec:stat_props} for which $n^c>50$. Calculating the asymmetry for each cluster, as opposed to for the entire system of particles, is necessary because multiple rotary worms may exist simultaneously. 

We then diagonalize the moment of inertia tensor so that it is of the form:
\begin{equation}
    \mathbf{I}=
\begin{bmatrix}
I_{1} & 0\\
0 & I_{2}
\end{bmatrix},
\end{equation}
where $I_{1}$ and $I_{2}$ are the principle moments of inertia. We characterize the asymmetry of the cluster as:
\begin{equation}
\label{eq:app_asymm}
    A=\frac{|I_1-I_2|}{I_1+I_2},
\end{equation}
such that $A=1$ means that the cluster is highly asymmetric and $A=0$ means that the cluster is circular. Equation~\eqref{eq:app_asymm} is shown as Eq.~\eqref{eq:asymm} in the main text.

\section{Stability analysis for the front of a worm}
\label{sec:app_frontworms}
To explore the behavior of a particle at the front of a worm, we now model the worm as a uniformly dense isosceles triangle (density $\rho$), which extends infinitely behind the particle of interest (particle $i$), which is at its front. All particles within the uniform, triangular sheet are oriented along the x-axis, such that $\phi_j=0$ for all $j$. The vertex angle of the triangle is $2\alpha$. The other two angles of the triangle are equal, but irrelevant for the following calculation. Particle $i$ is situated at the front of the worm, on the vertex angle as shown in Fig.~\ref{fig:front_worm}, and has orientation $\mathbf{u}_i=(\cos\phi_i,\sin\phi_i)$. We would like to determine the orientation of this particle when it interacts with the triangular sheet via the torques given in Eq.~\eqref{eq:prey-prey_aligngroup}.

\begin{figure}
  \centering
  \includegraphics[width=0.85\linewidth]{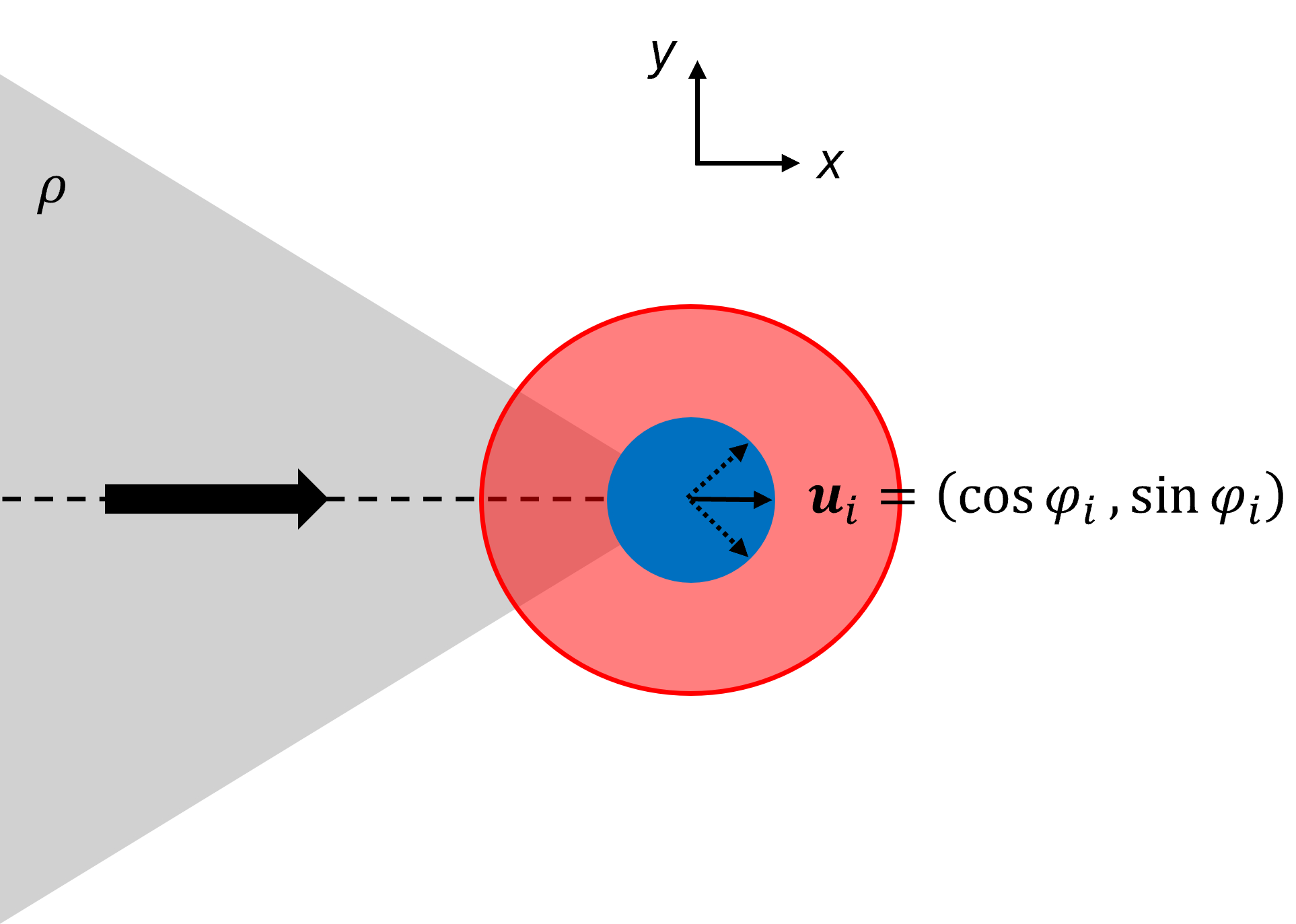}  
\caption{Schematic of particle at the front of a persistent worm, where the worm is approximated by an infinite triangular sheet of uniform density oriented along the x-axis. The particle of interest is shown in blue and the radius of interaction is represented by a red concentric circle around the particle.}
\label{fig:front_worm}
\end{figure}

Using Eq.~\eqref{eq:prey-predy_align} and the fact that $\phi_j=0$, we can write down the alignment torque on particle $i$ as:
\begin{align}
\label{eq:align_triangle}
\mathcal{T}^A_{ij}&=-\mathcal{T}_A\int^R_0\mathrm{d}r'\int^{\pi+\alpha}_{\pi-\alpha}\mathrm{d}\phi'~\rho r' \sin{(\phi_i)}\nonumber\\
&=-\rho R^2\alpha \mathcal{T}_A\sin{(\phi_i)}.
\end{align}
In Eq.~\eqref{eq:align_triangle}, the variables of integration, $r'$ and $\phi'$, represent the radial distance from particle $i$ and the angle from the x-axis respectively. Thus, the double integral in each equation goes over the area within the radius of interaction of particle $i$. The integral over $\phi'$ only goes from $\pi-\alpha$ to $\pi+\alpha$, in spite of the fact that no vision cone is used, because particles are only present within this range of angles.

Particle $i$ additionally experiences a cohesion torque (using Eq.~\eqref{eq:prey-prey_group}):
\begin{align}
\label{eq:cohesion_triangle}
\mathcal{T}^C_{ij}&=\mathcal{T}_C\int^R_0\mathrm{d}r'\int^{\pi+\alpha}_{\pi-\alpha}\mathrm{d}\phi'~\rho r' \Big(\cos{(\phi_i)}\sin{(\phi')} \nonumber \\
&\quad -\sin{(\phi_i)}\cos{(\phi')}\Big)\nonumber\\
&=\rho R^2\sin{(\alpha)}\mathcal{T}_C\sin{(\phi_i)}
\end{align}
The variables of integration in Eq.~\eqref{eq:cohesion_triangle} are the same as those in Eq.~\eqref{eq:align_triangle}.

We now let $\mathcal{T}_C=\mathcal{T}_0$. To correspond to our simulation results, we assume that $\mathcal{T}_A/\mathcal{T}_{C}=2$. Thus, we can write the total torque, $\mathcal{T}_{ij}$, on the particle as:
\begin{equation}
\label{eq:}
\mathcal{T}_{ij}=\rho R^2\sin{(\phi_i)}\mathcal{T}_0\Big(\sin{(\alpha)}-2\alpha\Big)
\end{equation}
The particle feels no torque when $\mathcal{T}_{ij}=0$, which occurs when the particle has an orientation $\phi_0=\pi n$, where $n$ is either $0$ or $1$ (theoretically it can be any integer, but physically it is limited to these two values). Theoretically, $\mathcal{T}_{ij}=0$ additionally when $\alpha=0$; however, this situation is not physical.

We now determine the stability of $\phi_0$ by looking at the behavior of $\mathcal{T}_{ij}$ near $\phi_0$,
\begin{equation}
\label{eq:}
\left.\frac{\partial \mathcal{T}_{ij}}{\partial \phi_i}\right\vert_{\phi_i=\phi_0}=\rho R^2\cos{(\phi_0)}\mathcal{T}_0\Big(\sin{(\alpha)}-2\alpha\Big)
\end{equation}
For $\phi_0=0$, $\left.\frac{\partial \mathcal{T}_{ij}}{\partial \phi_i}\right\vert_{\phi_i=\phi_0}<0$; therefore, this solution is stable. For $\phi_0=\pi$, $\left.\frac{\partial \mathcal{T}_{ij}}{\partial \phi_i}\right\vert_{\phi_i=\phi_0}>0$; therefore, this solution is unstable. The stability for each value of $\phi_0$ is independent of the value of $\alpha$: $\phi_0=0$ is stable for all values of $\alpha$ and $\phi_0=\pi$ is unstable for all values of alpha.


\end{appendices}


\bibliography{refs}


\begin{thebibliography}{53}
\ifx \bisbn   \undefined \def \bisbn  #1{ISBN #1}\fi
\ifx \binits  \undefined \def \binits#1{#1}\fi
\ifx \bauthor  \undefined \def \bauthor#1{#1}\fi
\ifx \batitle  \undefined \def \batitle#1{#1}\fi
\ifx \bjtitle  \undefined \def \bjtitle#1{#1}\fi
\ifx \bvolume  \undefined \def \bvolume#1{\textbf{#1}}\fi
\ifx \byear  \undefined \def \byear#1{#1}\fi
\ifx \bissue  \undefined \def \bissue#1{#1}\fi
\ifx \bfpage  \undefined \def \bfpage#1{#1}\fi
\ifx \blpage  \undefined \def \blpage #1{#1}\fi
\ifx \burl  \undefined \def \burl#1{\textsf{#1}}\fi
\ifx \doiurl  \undefined \def \doiurl#1{\url{https://doi.org/#1}}\fi
\ifx \betal  \undefined \def \betal{\textit{et al.}}\fi
\ifx \binstitute  \undefined \def \binstitute#1{#1}\fi
\ifx \binstitutionaled  \undefined \def \binstitutionaled#1{#1}\fi
\ifx \bctitle  \undefined \def \bctitle#1{#1}\fi
\ifx \beditor  \undefined \def \beditor#1{#1}\fi
\ifx \bpublisher  \undefined \def \bpublisher#1{#1}\fi
\ifx \bbtitle  \undefined \def \bbtitle#1{#1}\fi
\ifx \bedition  \undefined \def \bedition#1{#1}\fi
\ifx \bseriesno  \undefined \def \bseriesno#1{#1}\fi
\ifx \blocation  \undefined \def \blocation#1{#1}\fi
\ifx \bsertitle  \undefined \def \bsertitle#1{#1}\fi
\ifx \bsnm \undefined \def \bsnm#1{#1}\fi
\ifx \bsuffix \undefined \def \bsuffix#1{#1}\fi
\ifx \bparticle \undefined \def \bparticle#1{#1}\fi
\ifx \barticle \undefined \def \barticle#1{#1}\fi
\bibcommenthead
\ifx \bconfdate \undefined \def \bconfdate #1{#1}\fi
\ifx \botherref \undefined \def \botherref #1{#1}\fi
\ifx \url \undefined \def \url#1{\textsf{#1}}\fi
\ifx \bchapter \undefined \def \bchapter#1{#1}\fi
\ifx \bbook \undefined \def \bbook#1{#1}\fi
\ifx \bcomment \undefined \def \bcomment#1{#1}\fi
\ifx \oauthor \undefined \def \oauthor#1{#1}\fi
\ifx \citeauthoryear \undefined \def \citeauthoryear#1{#1}\fi
\ifx \endbibitem  \undefined \def \endbibitem {}\fi
\ifx \bconflocation  \undefined \def \bconflocation#1{#1}\fi
\ifx \arxivurl  \undefined \def \arxivurl#1{\textsf{#1}}\fi
\csname PreBibitemsHook\endcsname

\bibitem[\protect\citeauthoryear{Ben-Jacob et~al.}{1994}]{bacteria_patterns}
\begin{barticle}
\bauthor{\bsnm{Ben-Jacob}, \binits{E.}},
\bauthor{\bsnm{Shochet}, \binits{O.}},
\bauthor{\bsnm{Tenenbaum}, \binits{A.}},
\bauthor{\bsnm{Cohen}, \binits{I.}},
\bauthor{\bsnm{Czirók}, \binits{A.}},
\bauthor{\bsnm{Vicsek}, \binits{T.}}:
\batitle{Generic modeling of cooperative growth patterns in bacterial colonies}.
\bjtitle{Nature}
\bvolume{368},
\bfpage{46}--\blpage{49}
(\byear{1994})
\doiurl{10.1038/368046a0}
\end{barticle}
\endbibitem

\bibitem[\protect\citeauthoryear{Vicsek et~al.}{1990}]{bacteria_vicsek}
\begin{barticle}
\bauthor{\bsnm{Vicsek}, \binits{T.}},
\bauthor{\bsnm{Cserző}, \binits{M.}},
\bauthor{\bsnm{Horváth}, \binits{V.K.}}:
\batitle{Self-affine growth of bacterial colonies}.
\bjtitle{Physica A: Statistical Mechanics and its Applications}
\bvolume{167}(\bissue{2}),
\bfpage{315}--\blpage{321}
(\byear{1990})
\doiurl{10.1016/0378-4371(90)90116-A}
\end{barticle}
\endbibitem

\bibitem[\protect\citeauthoryear{Sokolov et~al.}{2007}]{bacteria3}
\begin{barticle}
\bauthor{\bsnm{Sokolov}, \binits{A.}},
\bauthor{\bsnm{Aranson}, \binits{I.S.}},
\bauthor{\bsnm{Kessler}, \binits{J.O.}},
\bauthor{\bsnm{Goldstein}, \binits{R.E.}}:
\batitle{Concentration dependence of the collective dynamics of swimming bacteria}.
\bjtitle{Phys. Rev. Lett.}
\bvolume{98},
\bfpage{158102}
(\byear{2007})
\doiurl{10.1103/PhysRevLett.98.158102}
\end{barticle}
\endbibitem

\bibitem[\protect\citeauthoryear{Cisneros et~al.}{2010}]{bacteria4}
\begin{botherref}
\oauthor{\bsnm{Cisneros}, \binits{L.}},
\oauthor{\bsnm{Cortez}, \binits{R.}},
\oauthor{\bsnm{Dombrowski}, \binits{C.}},
\oauthor{\bsnm{Goldstein}, \binits{R.E.}},
\oauthor{\bsnm{Kessler}, \binits{J.}}:
Fluid dynamics of self-propelled microorganisms, from individuals to concentrated populations.
Animal Locomotion,
99--115
(2010)
\doiurl{10.1007/978-3-642-11633-9_10}
\end{botherref}
\endbibitem

\bibitem[\protect\citeauthoryear{Schötz et~al.}{2008}]{zebrafish}
\begin{barticle}
\bauthor{\bsnm{Schötz}, \binits{E.-M.}},
\bauthor{\bsnm{Burdine}, \binits{R.}},
\bauthor{\bsnm{Jülicher}, \binits{F.}},
\bauthor{\bsnm{Steinberg}, \binits{M.}},
\bauthor{\bsnm{Heisenberg}, \binits{C.-P.}},
\bauthor{\bsnm{Foty}, \binits{R.}}:
\batitle{Quantitative differences in tissue surface tension influence zebrafish germ layer positioning}.
\bjtitle{HFSP journal}
\bvolume{2},
\bfpage{42}--\blpage{56}
(\byear{2008})
\doiurl{10.2976/1.2834817}
\end{barticle}
\endbibitem

\bibitem[\protect\citeauthoryear{Haas and Gilmour}{2006}]{tissue}
\begin{barticle}
\bauthor{\bsnm{Haas}, \binits{P.}},
\bauthor{\bsnm{Gilmour}, \binits{D.}}:
\batitle{Chemokine signaling mediates self-organizing tissue migration in the zebrafish lateral line}.
\bjtitle{Developmental Cell}
\bvolume{10}(\bissue{5}),
\bfpage{673}--\blpage{680}
(\byear{2006})
\doiurl{10.1016/j.devcel.2006.02.019}
\end{barticle}
\endbibitem

\bibitem[\protect\citeauthoryear{Friedl and Gilmour}{2009}]{morpho}
\begin{barticle}
\bauthor{\bsnm{Friedl}, \binits{P.}},
\bauthor{\bsnm{Gilmour}, \binits{D.}}:
\batitle{Collective cell migration in morphogenesis, regeneration and cancer}.
\bjtitle{Nature Reviews Molecular Cell Biology}
\bvolume{10},
\bfpage{445}--\blpage{457}
(\byear{2009})
\end{barticle}
\endbibitem

\bibitem[\protect\citeauthoryear{Ballerini et~al.}{2008}]{birds1}
\begin{barticle}
\bauthor{\bsnm{Ballerini}, \binits{M.}},
\bauthor{\bsnm{Cabibbo}, \binits{N.}},
\bauthor{\bsnm{Candelier}, \binits{R.}},
\bauthor{\bsnm{Cavagna}, \binits{A.}},
\bauthor{\bsnm{Cisbani}, \binits{E.}},
\bauthor{\bsnm{Giardina}, \binits{I.}},
\bauthor{\bsnm{Orlandi}, \binits{A.}},
\bauthor{\bsnm{Parisi}, \binits{G.}},
\bauthor{\bsnm{Procaccini}, \binits{A.}},
\bauthor{\bsnm{Viale}, \binits{M.}},
\bauthor{\bsnm{Zdravkovic}, \binits{V.}}:
\batitle{Empirical investigation of starling flocks: a benchmark study in collective animal behaviour}.
\bjtitle{Animal Behaviour}
\bvolume{76}(\bissue{1}),
\bfpage{201}--\blpage{215}
(\byear{2008})
\doiurl{10.1016/j.anbehav.2008.02.004}
\end{barticle}
\endbibitem

\bibitem[\protect\citeauthoryear{Cavagna and Giardina}{2014}]{birdworm1}
\begin{barticle}
\bauthor{\bsnm{Cavagna}, \binits{A.}},
\bauthor{\bsnm{Giardina}, \binits{I.}}:
\batitle{Bird flocks as condensed matter}.
\bjtitle{Annual Review of Condensed Matter Physics}
\bvolume{5}(\bissue{Volume 5, 2014}),
\bfpage{183}--\blpage{207}
(\byear{2014})
\doiurl{10.1146/annurev-conmatphys-031113-133834}
\end{barticle}
\endbibitem

\bibitem[\protect\citeauthoryear{Nagy et~al.}{2010}]{pigeons}
\begin{barticle}
\bauthor{\bsnm{Nagy}, \binits{M.}},
\bauthor{\bsnm{Akos}, \binits{Z.}},
\bauthor{\bsnm{Biro}, \binits{D.}},
\bauthor{\bsnm{Vicsek}, \binits{T.}}:
\batitle{Hierarchical group dynamics in pigeon flocks}.
\bjtitle{Nature}
\bvolume{464},
\bfpage{890}--\blpage{3}
(\byear{2010})
\doiurl{10.1038/nature08891}
\end{barticle}
\endbibitem

\bibitem[\protect\citeauthoryear{Hayakawa}{2010}]{birds2}
\begin{barticle}
\bauthor{\bsnm{Hayakawa}, \binits{Y.}}:
\batitle{Spatiotemporal dynamics of skeins of wild geese}.
\bjtitle{Europhysics Letters}
\bvolume{89}(\bissue{4}),
\bfpage{48004}
(\byear{2010})
\doiurl{10.1209/0295-5075/89/48004}
\end{barticle}
\endbibitem

\bibitem[\protect\citeauthoryear{Partridge et~al.}{1980}]{fish1}
\begin{barticle}
\bauthor{\bsnm{Partridge}, \binits{B.L.}},
\bauthor{\bsnm{Pitcher}, \binits{T.}},
\bauthor{\bsnm{Cullen}, \binits{J.M.}},
\bauthor{\bsnm{Wilson}, \binits{J.}}:
\batitle{The three-dimensional structure of fish schools}.
\bjtitle{Behavioral Ecology and Sociobiology}
\bvolume{6},
\bfpage{277}--\blpage{288}
(\byear{1980})
\end{barticle}
\endbibitem

\bibitem[\protect\citeauthoryear{Makris et~al.}{2009}]{herring}
\begin{barticle}
\bauthor{\bsnm{Makris}, \binits{N.C.}},
\bauthor{\bsnm{Ratilal}, \binits{P.}},
\bauthor{\bsnm{Jagannathan}, \binits{S.}},
\bauthor{\bsnm{Gong}, \binits{Z.}},
\bauthor{\bsnm{Andrews}, \binits{M.}},
\bauthor{\bsnm{Bertsatos}, \binits{I.}},
\bauthor{\bsnm{Godø}, \binits{O.R.}},
\bauthor{\bsnm{Nero}, \binits{R.W.}},
\bauthor{\bsnm{Jech}, \binits{J.M.}}:
\batitle{Critical population density triggers rapid formation of vast oceanic fish shoals}.
\bjtitle{Science}
\bvolume{323}(\bissue{5922}),
\bfpage{1734}--\blpage{1737}
(\byear{2009})
\doiurl{10.1126/science.1169441}
\end{barticle}
\endbibitem

\bibitem[\protect\citeauthoryear{Ward et~al.}{2008}]{fish3}
\begin{barticle}
\bauthor{\bsnm{Ward}, \binits{A.J.W.}},
\bauthor{\bsnm{Sumpter}, \binits{D.J.T.}},
\bauthor{\bsnm{Couzin}, \binits{I.D.}},
\bauthor{\bsnm{Hart}, \binits{P.J.B.}},
\bauthor{\bsnm{Krause}, \binits{J.}}:
\batitle{Quorum decision-making facilitates information transfer in fish shoals}.
\bjtitle{Proceedings of the National Academy of Sciences}
\bvolume{105}(\bissue{19}),
\bfpage{6948}--\blpage{6953}
(\byear{2008})
\doiurl{10.1073/pnas.0710344105}
\end{barticle}
\endbibitem

\bibitem[\protect\citeauthoryear{Reynolds}{1987}]{boid_model}
\begin{bchapter}
\bauthor{\bsnm{Reynolds}, \binits{C.W.}}:
\bctitle{Flocks, herds and schools: A distributed behavioral model}.
In: \bbtitle{Proceedings of the 14th Annual Conference on Computer Graphics and Interactive Techniques}.
\bsertitle{SIGGRAPH '87},
pp. \bfpage{25}--\blpage{34}.
\bpublisher{Association for Computing Machinery},
\blocation{New York, NY, USA}
(\byear{1987}).
\doiurl{10.1145/37401.37406}
\end{bchapter}
\endbibitem

\bibitem[\protect\citeauthoryear{Couzin et~al.}{2002}]{COUZIN20021}
\begin{barticle}
\bauthor{\bsnm{Couzin}, \binits{I.D.}},
\bauthor{\bsnm{Krause}, \binits{J.}},
\bauthor{\bsnm{James}, \binits{R.}},
\bauthor{\bsnm{Ruxton}, \binits{G.D.}},
\bauthor{\bsnm{Franks}, \binits{N.R.}}:
\batitle{Collective memory and spatial sorting in animal groups}.
\bjtitle{Journal of Theoretical Biology}
\bvolume{218}(\bissue{1}),
\bfpage{1}--\blpage{11}
(\byear{2002})
\doiurl{10.1006/jtbi.2002.3065}
\end{barticle}
\endbibitem

\bibitem[\protect\citeauthoryear{Couzin et~al.}{2005}]{Couzin_2005}
\begin{barticle}
\bauthor{\bsnm{Couzin}, \binits{I.}},
\bauthor{\bsnm{Krause}, \binits{J.}},
\bauthor{\bsnm{Franks}, \binits{N.}},
\bauthor{\bsnm{Levin}, \binits{S.}}:
\batitle{Effective leadership and decision-making in animal groups on the move}.
\bjtitle{Nature}
\bvolume{433},
\bfpage{513}--\blpage{6}
(\byear{2005})
\doiurl{10.1038/nature03236}
\end{barticle}
\endbibitem

\bibitem[\protect\citeauthoryear{Vicsek et~al.}{1995}]{vicsek}
\begin{barticle}
\bauthor{\bsnm{Vicsek}, \binits{T.}},
\bauthor{\bsnm{Czir\'ok}, \binits{A.}},
\bauthor{\bsnm{Ben-Jacob}, \binits{E.}},
\bauthor{\bsnm{Cohen}, \binits{I.}},
\bauthor{\bsnm{Shochet}, \binits{O.}}:
\batitle{Novel type of phase transition in a system of self-driven particles}.
\bjtitle{Phys. Rev. Lett.}
\bvolume{75},
\bfpage{1226}--\blpage{1229}
(\byear{1995})
\doiurl{10.1103/PhysRevLett.75.1226}
\end{barticle}
\endbibitem

\bibitem[\protect\citeauthoryear{Romanczuk and Schimansky-Geier}{2012}]{ROMANCZUK}
\begin{barticle}
\bauthor{\bsnm{Romanczuk}, \binits{P.}},
\bauthor{\bsnm{Schimansky-Geier}, \binits{L.}}:
\batitle{Mean-field theory of collective motion due to velocity alignment}.
\bjtitle{Ecological Complexity}
\bvolume{10},
\bfpage{83}--\blpage{92}
(\byear{2012})
\doiurl{10.1016/j.ecocom.2011.07.008} .
\bcomment{From spatially explicit population models to mean-field dynamics}
\end{barticle}
\endbibitem

\bibitem[\protect\citeauthoryear{Martín-Gómez et~al.}{2018}]{Pagonabarraga}
\begin{barticle}
\bauthor{\bsnm{Martín-Gómez}, \binits{A.}},
\bauthor{\bsnm{Levis}, \binits{D.}},
\bauthor{\bsnm{Díaz-Guilera}, \binits{A.}},
\bauthor{\bsnm{Pagonabarraga}, \binits{I.}}:
\batitle{Collective motion of active brownian particles with polar alignment}.
\bjtitle{Soft Matter}
\bvolume{14},
\bfpage{2610}--\blpage{2618}
(\byear{2018})
\doiurl{10.1039/C8SM00020D}
\end{barticle}
\endbibitem

\bibitem[\protect\citeauthoryear{Caprini and L\"owen}{2023}]{Loewen}
\begin{barticle}
\bauthor{\bsnm{Caprini}, \binits{L.}},
\bauthor{\bsnm{L\"owen}, \binits{H.}}:
\batitle{Flocking without alignment interactions in attractive active brownian particles}.
\bjtitle{Phys. Rev. Lett.}
\bvolume{130},
\bfpage{148202}
(\byear{2023})
\doiurl{10.1103/PhysRevLett.130.148202}
\end{barticle}
\endbibitem

\bibitem[\protect\citeauthoryear{Barberis and Peruani}{2016}]{Peruani_attraction}
\begin{barticle}
\bauthor{\bsnm{Barberis}, \binits{L.}},
\bauthor{\bsnm{Peruani}, \binits{F.}}:
\batitle{Large-scale patterns in a minimal cognitive flocking model: Incidental leaders, nematic patterns, and aggregates}.
\bjtitle{Phys. Rev. Lett.}
\bvolume{117},
\bfpage{248001}
(\byear{2016})
\doiurl{10.1103/PhysRevLett.117.248001}
\end{barticle}
\endbibitem

\bibitem[\protect\citeauthoryear{Grégoire et~al.}{2003}]{vicsek+cohesion}
\begin{barticle}
\bauthor{\bsnm{Grégoire}, \binits{G.}},
\bauthor{\bsnm{Chaté}, \binits{H.}},
\bauthor{\bsnm{Tu}, \binits{Y.}}:
\batitle{Moving and staying together without a leader}.
\bjtitle{Physica D: Nonlinear Phenomena}
\bvolume{181}(\bissue{3}),
\bfpage{157}--\blpage{170}
(\byear{2003})
\doiurl{10.1016/S0167-2789(03)00102-7}
\end{barticle}
\endbibitem

\bibitem[\protect\citeauthoryear{Ivlev et~al.}{2015}]{statmech_nr}
\begin{barticle}
\bauthor{\bsnm{Ivlev}, \binits{A.V.}},
\bauthor{\bsnm{Bartnick}, \binits{J.}},
\bauthor{\bsnm{Heinen}, \binits{M.}},
\bauthor{\bsnm{Du}, \binits{C.-R.}},
\bauthor{\bsnm{Nosenko}, \binits{V.}},
\bauthor{\bsnm{L\"owen}, \binits{H.}}:
\batitle{Statistical mechanics where newton's third law is broken}.
\bjtitle{Phys. Rev. X}
\bvolume{5},
\bfpage{011035}
(\byear{2015})
\doiurl{10.1103/PhysRevX.5.011035}
\end{barticle}
\endbibitem

\bibitem[\protect\citeauthoryear{Osat and Golestanian}{2022}]{Golestanian_nr}
\begin{barticle}
\bauthor{\bsnm{Osat}, \binits{S.}},
\bauthor{\bsnm{Golestanian}, \binits{R.}}:
\batitle{Non-reciprocal multifarious self-organization}.
\bjtitle{Nature Nanotechnology}
\bvolume{18},
\bfpage{1}--\blpage{7}
(\byear{2022})
\doiurl{10.1038/s41565-022-01258-2}
\end{barticle}
\endbibitem

\bibitem[\protect\citeauthoryear{{Klapp}}{2023}]{sabine_nr}
\begin{barticle}
\bauthor{\bsnm{{Klapp}}, \binits{S.H.L.}}:
\batitle{{Non-reciprocal interaction for living matter}}.
\bjtitle{Nature Nanotechnology}
\bvolume{18}(\bissue{1}),
\bfpage{8}--\blpage{9}
(\byear{2023})
\doiurl{10.1038/s41565-022-01268-0}
\end{barticle}
\endbibitem

\bibitem[\protect\citeauthoryear{Kreienkamp and Klapp}{2022}]{Kreienkamp_2022}
\begin{barticle}
\bauthor{\bsnm{Kreienkamp}, \binits{K.L.}},
\bauthor{\bsnm{Klapp}, \binits{S.H.L.}}:
\batitle{Clustering and flocking of repulsive chiral active particles with non-reciprocal couplings}.
\bjtitle{New Journal of Physics}
\bvolume{24}(\bissue{12}),
\bfpage{123009}
(\byear{2022})
\doiurl{10.1088/1367-2630/ac9cc3}
\end{barticle}
\endbibitem

\bibitem[\protect\citeauthoryear{Knežević et~al.}{2022}]{Milos_Nonreciprocal}
\begin{botherref}
\oauthor{\bsnm{Knežević}, \binits{M.}},
\oauthor{\bsnm{Welker}, \binits{T.}},
\oauthor{\bsnm{Stark}, \binits{H.}}:
Collective motion of active particles exhibiting non-reciprocal orientational interactions.
Scientific Reports
\textbf{12}
(2022)
\doiurl{10.1038/s41598-022-23597-9}
\end{botherref}
\endbibitem

\bibitem[\protect\citeauthoryear{Das et~al.}{2024}]{Alert_TurnAway}
\begin{barticle}
\bauthor{\bsnm{Das}, \binits{S.}},
\bauthor{\bsnm{Ciarchi}, \binits{M.}},
\bauthor{\bsnm{Zhou}, \binits{Z.}},
\bauthor{\bsnm{Yan}, \binits{J.}},
\bauthor{\bsnm{Zhang}, \binits{J.}},
\bauthor{\bsnm{Alert}, \binits{R.}}:
\batitle{Flocking by turning away}.
\bjtitle{Phys. Rev. X}
\bvolume{14},
\bfpage{031008}
(\byear{2024})
\doiurl{10.1103/PhysRevX.14.031008}
\end{barticle}
\endbibitem

\bibitem[\protect\citeauthoryear{Nilsson and Volpe}{2017}]{Nilsson_2017}
\begin{barticle}
\bauthor{\bsnm{Nilsson}, \binits{S.}},
\bauthor{\bsnm{Volpe}, \binits{G.}}:
\batitle{Metastable clusters and channels formed by active particles with aligning interactions}.
\bjtitle{New Journal of Physics}
\bvolume{19}(\bissue{11}),
\bfpage{115008}
(\byear{2017})
\doiurl{10.1088/1367-2630/aa9516}
\end{barticle}
\endbibitem

\bibitem[\protect\citeauthoryear{Zhang et~al.}{2021}]{Alert_TurnTowards}
\begin{botherref}
\oauthor{\bsnm{Zhang}, \binits{J.}},
\oauthor{\bsnm{Alert}, \binits{R.}},
\oauthor{\bsnm{Yan}, \binits{J.}},
\oauthor{\bsnm{Wingreen}, \binits{N.}},
\oauthor{\bsnm{Granick}, \binits{S.}}:
Active phase separation by turning towards regions of higher density.
Nature Physics
\textbf{17}
(2021)
\doiurl{10.1038/s41567-021-01238-8}
\end{botherref}
\endbibitem

\bibitem[\protect\citeauthoryear{Fruchart et~al.}{2021}]{Fruchart2021NonreciprocalPT}
\begin{barticle}
\bauthor{\bsnm{Fruchart}, \binits{M.}},
\bauthor{\bsnm{Hanai}, \binits{R.}},
\bauthor{\bsnm{Littlewood}, \binits{P.B.}},
\bauthor{\bsnm{Vitelli}, \binits{V.}}:
\batitle{Non-reciprocal phase transitions}.
\bjtitle{Nature}
\bvolume{592},
\bfpage{363}--\blpage{369}
(\byear{2021})
\end{barticle}
\endbibitem

\bibitem[\protect\citeauthoryear{You et~al.}{2020}]{Marchetti_nr}
\begin{barticle}
\bauthor{\bsnm{You}, \binits{Z.}},
\bauthor{\bsnm{Baskaran}, \binits{A.}},
\bauthor{\bsnm{Marchetti}, \binits{M.C.}}:
\batitle{Nonreciprocity as a generic route to traveling states}.
\bjtitle{Proceedings of the National Academy of Sciences}
\bvolume{117}(\bissue{33}),
\bfpage{19767}--\blpage{19772}
(\byear{2020})
\doiurl{10.1073/pnas.2010318117}
\end{barticle}
\endbibitem

\bibitem[\protect\citeauthoryear{Saha et~al.}{2019}]{Saha_2019}
\begin{barticle}
\bauthor{\bsnm{Saha}, \binits{S.}},
\bauthor{\bsnm{Ramaswamy}, \binits{S.}},
\bauthor{\bsnm{Golestanian}, \binits{R.}}:
\batitle{Pairing, waltzing and scattering of chemotactic active colloids}.
\bjtitle{New Journal of Physics}
\bvolume{21}(\bissue{6}),
\bfpage{063006}
(\byear{2019})
\doiurl{10.1088/1367-2630/ab20fd}
\end{barticle}
\endbibitem

\bibitem[\protect\citeauthoryear{Zhang and Garcia-Millan}{2023}]{Rosalba}
\begin{barticle}
\bauthor{\bsnm{Zhang}, \binits{Z.}},
\bauthor{\bsnm{Garcia-Millan}, \binits{R.}}:
\batitle{Entropy production of nonreciprocal interactions}.
\bjtitle{Phys. Rev. Res.}
\bvolume{5},
\bfpage{022033}
(\byear{2023})
\doiurl{10.1103/PhysRevResearch.5.L022033}
\end{barticle}
\endbibitem

\bibitem[\protect\citeauthoryear{Bhattacherjee et~al.}{2024}]{NR_force}
\begin{barticle}
\bauthor{\bsnm{Bhattacherjee}, \binits{B.}},
\bauthor{\bsnm{Hayakawa}, \binits{M.}},
\bauthor{\bsnm{Shibata}, \binits{T.}}:
\batitle{Structure formation induced by non-reciprocal cell–cell interactions in a multicellular system}.
\bjtitle{Soft Matter}
\bvolume{20},
\bfpage{2739}--\blpage{2749}
(\byear{2024})
\doiurl{10.1039/D3SM01752D}
\end{barticle}
\endbibitem

\bibitem[\protect\citeauthoryear{Negi et~al.}{2024}]{Gompper_similar}
\begin{barticle}
\bauthor{\bsnm{Negi}, \binits{R.S.}},
\bauthor{\bsnm{Winkler}, \binits{R.G.}},
\bauthor{\bsnm{Gompper}, \binits{G.}}:
\batitle{Collective behavior of self-steering active particles with velocity alignment and visual perception}.
\bjtitle{Phys. Rev. Res.}
\bvolume{6},
\bfpage{013118}
(\byear{2024})
\doiurl{10.1103/PhysRevResearch.6.013118}
\end{barticle}
\endbibitem

\bibitem[\protect\citeauthoryear{{Weeks} et~al.}{1971}]{WCA}
\begin{barticle}
\bauthor{\bsnm{{Weeks}}, \binits{J.D.}},
\bauthor{\bsnm{{Chandler}}, \binits{D.}},
\bauthor{\bsnm{{Andersen}}, \binits{H.C.}}:
\batitle{{Role of Repulsive Forces in Determining the Equilibrium Structure of Simple Liquids}}.
\bjtitle{J. Phys. Chem.}
\bvolume{54}(\bissue{12}),
\bfpage{5237}--\blpage{5247}
(\byear{1971})
\doiurl{10.1063/1.1674820}
\end{barticle}
\endbibitem

\bibitem[\protect\citeauthoryear{Kryuchkov et~al.}{2024}]{Kryuchkov}
\begin{barticle}
\bauthor{\bsnm{Kryuchkov}, \binits{N.P.}},
\bauthor{\bsnm{Nasyrov}, \binits{A.D.}},
\bauthor{\bsnm{Gursky}, \binits{K.D.}},
\bauthor{\bsnm{Yurchenko}, \binits{S.O.}}:
\batitle{Influence of anomalous agents on the dynamics of an active system}.
\bjtitle{Phys. Rev. E}
\bvolume{109},
\bfpage{034601}
(\byear{2024})
\doiurl{10.1103/PhysRevE.109.034601}
\end{barticle}
\endbibitem

\bibitem[\protect\citeauthoryear{Kryuchkov et~al.}{2020}]{Kryuchkov2}
\begin{barticle}
\bauthor{\bsnm{Kryuchkov}, \binits{N.P.}},
\bauthor{\bsnm{Mistryukova}, \binits{L.A.}},
\bauthor{\bsnm{Sapelkin}, \binits{A.V.}},
\bauthor{\bsnm{Yurchenko}, \binits{S.O.}}:
\batitle{Strange attractors induced by melting in systems with nonreciprocal effective interactions}.
\bjtitle{Phys. Rev. E}
\bvolume{101},
\bfpage{063205}
(\byear{2020})
\doiurl{10.1103/PhysRevE.101.063205}
\end{barticle}
\endbibitem

\bibitem[\protect\citeauthoryear{Kryuchkov et~al.}{2018}]{Kryuchkov3}
\begin{barticle}
\bauthor{\bsnm{Kryuchkov}, \binits{N.P.}},
\bauthor{\bsnm{Ivlev}, \binits{A.V.}},
\bauthor{\bsnm{Yurchenko}, \binits{S.O.}}:
\batitle{Dissipative phase transitions in systems with nonreciprocal effective interactions}.
\bjtitle{Soft Matter}
\bvolume{14},
\bfpage{9720}--\blpage{9729}
(\byear{2018})
\doiurl{10.1039/C8SM01836G}
\end{barticle}
\endbibitem

\bibitem[\protect\citeauthoryear{Gogia and Burton}{2017}]{Gogia}
\begin{barticle}
\bauthor{\bsnm{Gogia}, \binits{G.}},
\bauthor{\bsnm{Burton}, \binits{J.C.}}:
\batitle{Emergent bistability and switching in a nonequilibrium crystal}.
\bjtitle{Phys. Rev. Lett.}
\bvolume{119},
\bfpage{178004}
(\byear{2017})
\doiurl{10.1103/PhysRevLett.119.178004}
\end{barticle}
\endbibitem

\bibitem[\protect\citeauthoryear{Ester et~al.}{1996}]{dbscan1}
\begin{bchapter}
\bauthor{\bsnm{Ester}, \binits{M.}},
\bauthor{\bsnm{Kriegel}, \binits{H.-P.}},
\bauthor{\bsnm{Sander}, \binits{J.}},
\bauthor{\bsnm{Xu}, \binits{X.}}:
\bctitle{A density-based algorithm for discovering clusters in large spatial databases with noise}.
In: \bbtitle{Knowledge Discovery and Data Mining}
(\byear{1996})
\end{bchapter}
\endbibitem

\bibitem[\protect\citeauthoryear{Schubert et~al.}{2017}]{dbscan2}
\begin{botherref}
\oauthor{\bsnm{Schubert}, \binits{E.}},
\oauthor{\bsnm{Sander}, \binits{J.}},
\oauthor{\bsnm{Ester}, \binits{M.}},
\oauthor{\bsnm{Kriegel}, \binits{H.P.}},
\oauthor{\bsnm{Xu}, \binits{X.}}:
Dbscan revisited, revisited: Why and how you should (still) use dbscan.
ACM Trans. Database Syst.
\textbf{42}(3)
(2017)
\doiurl{10.1145/3068335}
\end{botherref}
\endbibitem

\bibitem[\protect\citeauthoryear{Pedregosa et~al.}{2011}]{scikit-learn}
\begin{barticle}
\bauthor{\bsnm{Pedregosa}, \binits{F.}},
\bauthor{\bsnm{Varoquaux}, \binits{G.}},
\bauthor{\bsnm{Gramfort}, \binits{A.}},
\bauthor{\bsnm{Michel}, \binits{V.}},
\bauthor{\bsnm{Thirion}, \binits{B.}},
\bauthor{\bsnm{Grisel}, \binits{O.}},
\bauthor{\bsnm{Blondel}, \binits{M.}},
\bauthor{\bsnm{Prettenhofer}, \binits{P.}},
\bauthor{\bsnm{Weiss}, \binits{R.}},
\bauthor{\bsnm{Dubourg}, \binits{V.}},
\bauthor{\bsnm{Vanderplas}, \binits{J.}},
\bauthor{\bsnm{Passos}, \binits{A.}},
\bauthor{\bsnm{Cournapeau}, \binits{D.}},
\bauthor{\bsnm{Brucher}, \binits{M.}},
\bauthor{\bsnm{Perrot}, \binits{M.}},
\bauthor{\bsnm{Duchesnay}, \binits{E.}}:
\batitle{Scikit-learn: Machine learning in {P}ython}.
\bjtitle{Journal of Machine Learning Research}
\bvolume{12},
\bfpage{2825}--\blpage{2830}
(\byear{2011})
\end{barticle}
\endbibitem

\bibitem[\protect\citeauthoryear{Bai and Breen}{2008}]{Bai_Breen_pbc}
\begin{barticle}
\bauthor{\bsnm{Bai}, \binits{L.}},
\bauthor{\bsnm{Breen}, \binits{D.}}:
\batitle{Calculating center of mass in an unbounded 2d environment}.
\bjtitle{Journal of Graphics Tools}
\bvolume{13}(\bissue{4}),
\bfpage{53}--\blpage{60}
(\byear{2008})
\doiurl{10.1080/2151237X.2008.10129266}
\end{barticle}
\endbibitem

\bibitem[\protect\citeauthoryear{Isele-Holder et~al.}{2015}]{spiral_activefilament}
\begin{barticle}
\bauthor{\bsnm{Isele-Holder}, \binits{R.E.}},
\bauthor{\bsnm{Elgeti}, \binits{J.}},
\bauthor{\bsnm{Gompper}, \binits{G.}}:
\batitle{Self-propelled worm-like filaments: spontaneous spiral formation{,} structure{,} and dynamics}.
\bjtitle{Soft Matter}
\bvolume{11},
\bfpage{7181}--\blpage{7190}
(\byear{2015})
\doiurl{10.1039/C5SM01683E}
\end{barticle}
\endbibitem

\bibitem[\protect\citeauthoryear{Cavagna et~al.}{2013}]{change_neighbors}
\begin{barticle}
\bauthor{\bsnm{Cavagna}, \binits{A.}},
\bauthor{\bsnm{Queir{\'o}s}, \binits{S.D.}},
\bauthor{\bsnm{Giardina}, \binits{I.}},
\bauthor{\bsnm{Stefanini}, \binits{F.}},
\bauthor{\bsnm{Viale}, \binits{M.}}:
\batitle{Diffusion of individual birds in starling flocks}.
\bjtitle{Proceedings of the Royal Society B: Biological Sciences}
\bvolume{280}(\bissue{1756}),
\bfpage{20122484}
(\byear{2013})
\end{barticle}
\endbibitem

\bibitem[\protect\citeauthoryear{Voelkl et~al.}{2015}]{bird_leaders}
\begin{barticle}
\bauthor{\bsnm{Voelkl}, \binits{B.}},
\bauthor{\bsnm{Portugal}, \binits{S.J.}},
\bauthor{\bsnm{Unsöld}, \binits{M.}},
\bauthor{\bsnm{Usherwood}, \binits{J.R.}},
\bauthor{\bsnm{Wilson}, \binits{A.M.}},
\bauthor{\bsnm{Fritz}, \binits{J.}}:
\batitle{Matching times of leading and following suggest cooperation through direct reciprocity during v-formation flight in ibis}.
\bjtitle{Proceedings of the National Academy of Sciences}
\bvolume{112}(\bissue{7}),
\bfpage{2115}--\blpage{2120}
(\byear{2015})
\doiurl{10.1073/pnas.1413589112}
\end{barticle}
\endbibitem

\bibitem[\protect\citeauthoryear{Dormann and Weijer}{2001}]{slug_cells}
\begin{barticle}
\bauthor{\bsnm{Dormann}, \binits{D.}},
\bauthor{\bsnm{Weijer}, \binits{C.J.}}:
\batitle{Propagating chemoattractant waves coordinate periodic cell movement in dictyostelium slugs}.
\bjtitle{Development}
\bvolume{128}(\bissue{22}),
\bfpage{4535}--\blpage{4543}
(\byear{2001})
\doiurl{10.1242/dev.128.22.4535}
\end{barticle}
\endbibitem

\bibitem[\protect\citeauthoryear{Palsson and Othmer}{2000}]{slugs_cells2}
\begin{barticle}
\bauthor{\bsnm{Palsson}, \binits{E.}},
\bauthor{\bsnm{Othmer}, \binits{H.G.}}:
\batitle{A model for individual and collective cell movement in \textit{Dictyostelium discoideum}}.
\bjtitle{Proceedings of the National Academy of Sciences}
\bvolume{97}(\bissue{19}),
\bfpage{10448}--\blpage{10453}
(\byear{2000})
\doiurl{10.1073/pnas.97.19.10448}
\end{barticle}
\endbibitem

\bibitem[\protect\citeauthoryear{Chen and Bechinger}{2022}]{microrobots1}
\begin{botherref}
\oauthor{\bsnm{Chen}, \binits{C.-J.}},
\oauthor{\bsnm{Bechinger}, \binits{C.}}:
Collective response of microrobotic swarms to external threats.
New Journal of Physics
\textbf{24}
(2022)
\doiurl{10.1088/1367-2630/ac5374}
\end{botherref}
\endbibitem

\bibitem[\protect\citeauthoryear{Heuthe et~al.}{2024}]{microrobots2}
\begin{barticle}
\bauthor{\bsnm{Heuthe}, \binits{V.-L.}},
\bauthor{\bsnm{Panizon}, \binits{E.}},
\bauthor{\bsnm{Gu}, \binits{H.}},
\bauthor{\bsnm{Bechinger}, \binits{C.}}:
\batitle{Counterfactual rewards promote collective transport using individually controlled swarm microrobots}.
\bjtitle{Science Robotics}
\bvolume{9}(\bissue{97}),
\bfpage{5888}
(\byear{2024})
\doiurl{10.1126/scirobotics.ado5888}
\end{barticle}
\endbibitem

\end{thebibliography}

\end{document}